\documentclass[lettersize,journal]{IEEEtran}
\usepackage{amsmath,amsfonts}
\usepackage{algorithmic}
\usepackage{algorithm}
\usepackage{array}
\usepackage[caption=false,font=normalsize,labelfont=sf,textfont=sf]{subfig}
\usepackage{textcomp}
\usepackage{stfloats}
\usepackage{url}
\usepackage{verbatim}
\usepackage{graphicx}
\usepackage{cite}
\usepackage{makecell}
\usepackage{booktabs}
\usepackage{flushend}
\usepackage{color}
\usepackage{multirow}
\usepackage{indentfirst}
\usepackage{float}
\usepackage{microtype}
\usepackage{colortbl} 
\usepackage[]{listings}
\usepackage{xcolor}
\usepackage{xspace}

\definecolor{codegreen}{rgb}{0,0.6,0}
\definecolor{codegray}{rgb}{0.5,0.5,0.5}
\definecolor{codepurple}{rgb}{0.58,0,0.82}
\definecolor{shallowred}{rgb}{1,0.8,0.8}

\lstdefinestyle{mystyle}{
    language=Java,
    commentstyle=\color{codegreen},
    keywordstyle=\color{blue},
    stringstyle=\color{codepurple},
    basicstyle=\fontfamily{zi4}\selectfont\scriptsize,
    breakatwhitespace=false,
    breaklines=true,                 
    captionpos=b,                    
    keepspaces=true,                 
    numbers=left,                    
    numbersep=5pt,                  
    showspaces=false,                
    showstringspaces=false,
    frame=single,                    
    framexleftmargin=0mm,            
    framexrightmargin=0mm,           
    framextopmargin=0mm,             
    framexbottommargin=0mm,    
    showtabs=false,                  
    tabsize=2,
    escapeinside={(*@}{@*)}, 
}
\usepackage{bbding}
\usepackage{pifont}
\usepackage{hyperref}
\usepackage{cleveref}
\newcommand{\jy}[1]{\textcolor{magenta}{#1}}
\newcommand{\tool}{\textit{VulsTotal}\xspace}

\begin{document}
\title{A Comprehensive Study on Static Application Security Testing (SAST) Tools for Android}

\author{Jingyun Zhu, Kaixuan Li, Sen Chen, Lingling Fan, Junjie Wang, and Xiaofei Xie
\thanks{Jingyun Zhu and Kaixuan Li contributed equally to this work.}
\thanks{Jingyun Zhu, Sen Chen (Corresponding author), and Junjie Wang are with the College of Intelligence and Computing, Tianjin University, China, 300350.}
\thanks{Kaixuan Li is with the East China Normal University, China.}

\thanks{Lingling Fan is with the Nankai University, China.}

\thanks{Xiaofei Xie is with the Singapore Management University, Singapore.}
}

\markboth{Journal of \LaTeX\ Class Files,~Vol.~14, No.~8, August~2021}
{Shell \MakeLowercase{\textit{et al.}}: A Sample Article Using IEEEtran.cls for IEEE Journals}
\IEEEpubid{0000--0000/00\$00.00~\copyright~2024 IEEE}
\maketitle

\begin{abstract}
To identify security vulnerabilities in Android applications, numerous static application security testing (SAST) tools have been proposed. However, it poses significant challenges to assess their overall performance on diverse vulnerability types. The task is non-trivial and poses considerable challenges. 
{Firstly, the absence of a unified evaluation platform for defining and describing tools' supported vulnerability types, coupled with the lack of normalization for the intricate and varied reports generated by different tools, significantly adds to the complexity.}
Secondly, there is a scarcity of adequate benchmarks, particularly those derived from real-world scenarios.
To address these problems, we are the first to propose a unified platform named \tool, {supporting various vulnerability types, enabling comprehensive and versatile analysis across diverse SAST tools}. 
Specifically, we begin by meticulously selecting 11 free and open-sourced SAST tools from a pool of 97 existing options, adhering to clearly defined criteria. After that, we invest significant efforts in comprehending the detection rules of each tool, subsequently unifying 67 general/common vulnerability types for {Android} SAST tools. We also redefine and implement a standardized reporting format, ensuring uniformity in presenting results across all tools.
Additionally, to mitigate the problem of benchmarks, we conducted a manual analysis of huge amounts of CVEs to construct a new CVE-based benchmark based on our comprehension of Android app vulnerabilities. 
Leveraging the evaluation platform, which integrates both existing synthetic benchmarks and newly constructed CVE-based benchmarks from this study, we conducted a comprehensive analysis to evaluate and compare these selected tools from various perspectives, such as general vulnerability type coverage, type consistency, tool effectiveness, and time performance. Our observations yielded impressive findings, like the technical reasons underlying the performance, which provide insights for different stakeholders.
\end{abstract}


\vspace{50pt}

\section{Introduction}
\IEEEPARstart{R}{ecently}, mobile devices have become an indispensable part of people's daily lives. They serve as a platform for numerous mobile applications (apps) catering to various needs, such as shopping, banking, and music, among others~\cite{chen2019storydroid, chen2022automatically,zhang2023web}. 
While these apps greatly enhance convenience, they also store a vast amount of user-related information, leading to security risks such as sensitive data leakage~\cite{arzt2014flowdroid,li2015iccta,wei2018amandroid,chen2020empirical} and ACE attack~\cite{mitreCVE20193568}.
{For example, a critical zero-day vulnerability~\cite{mitreCVE20193568} discovered in WhatsApp allows attackers to remotely install spyware via specially crafted SRTCP packets. Exploited by NSO Group, it executed arbitrary code without requiring user call response, impacting numerous users.}
Consequently, guaranteeing the safety and dependability of mobile apps has become a top priority for all stakeholders. To ensure the reliability of mobile apps, both academia and industry have made significant efforts. A plethora of Static {Application} Security Tools (SAST) for checking security vulnerabilities have been developed. These tools play a vital role in identifying potential threats and mitigating security risks, thus enhancing the overall security posture of mobile apps~\cite{AndroBugs, QARK, SUPER, SPECK,chen2022ausera,JAADAS,MobSF,Marvin}.
\IEEEpubidadjcol

Evaluating the overall effectiveness of SAST tools offers significant benefits to various stakeholders, including tool developers, users, and researchers. 
While numerous tools have been designed to address specific vulnerability types, it is crucial to grasp how well SAST tools work with general vulnerability types. This understanding serves as a guidepost for developers to bolster support for various general or common vulnerability types in the Android domain and also aids users in choosing tools offering broader, more inclusive vulnerability detection.
The existing studies~\cite{zhang2021analyzing,ranganath2020free,senanayake2023android} have been conducted to evaluate the detection capabilities, but they often suffer from two main problems. 
\textbf{\textit{(1)}} Firstly, their absence of a unified platform means that comparisons can only focus on coarse-grained quantities rather than fine-grained vulnerability types. 
{For instance, the Android SAST tool named SUPER~\cite{SUPER} consolidates various cryptographic vulnerability types under the broad type of ``Weak Algorithms'' whereas other tools, like AUSERA~\cite{chen2022ausera,chen2020empirical,DBLP:conf/sigsoft/ChenSFMXLX18}, offer a more detailed breakdown, distinguishing between ``AES encryption issue'' and ``DES encryption issue''. This leads existing approaches to prefer to evaluate these vulnerabilities primarily under the broad ``Cryptography'' category at a coarse-grained level.}
{Moreover, the lack of normalization across diverse tool reports also amplifies complexity.} 
These hinder a comprehensive understanding of the strengths and weaknesses of different tools from this important aspect, limiting their potential for further improvement.
\textbf{\textit{(2)}} Secondly, the evaluation process typically relies solely on synthetic benchmarks~\cite{mitra2019benchpress}, which may not precisely represent real-world scenarios. Hence, the effectiveness of these tools in real-world environments may not be adequately gauged, potentially leading to discrepancies between lab-based assessments and practical applications.

Indeed, conducting a comprehensive evaluation of SAST tools faces substantial challenges that need to be addressed to improve the evaluation process effectively. 
\textbf{\textit{(1)}} One of the significant obstacles is the various vulnerability types supported by different SAST tools, tailored to their specific detection scenarios. Consequently, direct comparisons of their supported types become impractical due to the lack of standardized documentation specifications for many tools. 
{As for the issue of varied report formats and contents among SAST tools, this creates barriers to directly comparing valuable vulnerability reports across tools.}
To overcome these, huge efforts should focus on establishing a unified platform or set of guidelines for defining and describing their complex and diverse vulnerability types, {and normalizing vulnerability reports format for enabling automatic comparison}, allowing for more meaningful and fair {evaluation} between different tools.
\textbf{\textit{(2)}} Further, as synthetic benchmarks are widely used in evaluating SAST tools, we endeavored to comprehensively evaluate the performance of these tools by constructing a real-world benchmark based on Android-specific CVEs. However, challenges arose due to the lack of clarity in the descriptions provided by some CVEs and the absence of detailed vulnerability information.

In detail, to tackle these challenges, we first selected 11 free and open-sourced Android SAST tools based on well-defined criteria from 99 existing static analysis tools as platform bases. We then meticulously reviewed the metadata of each SAST tool and unified the various supported vulnerability types of different tools, resulting in 67 unified {\textit{general/common} types within Android landscope} as a taxonomy.
{Further, we adjusted SAST tools' source code for unified TXT result reports and crafted parsers to extract vulnerability reports achieving normalization.}
Based on the tool bases, unified taxonomy, {and parsers}, we proposed a platform, named \tool, to help effectively evaluate the detection capability of Android SAST tools. We highlight the aforementioned key steps in developing the platform required a total investment of five person-months.
Secondly, to overcome the challenges of constructing real-world benchmarks, we initially employed automated methods to filter out-of-scope CVEs. Subsequently, we dedicated significant human effort to manually label the remaining CVEs based on their descriptions and provided resources. This meticulous process allowed us to build a CVE-based benchmark tailored to our research scope.
We utilized the platform and performed a comprehensive evaluation of selected SAST tools based on different synthetic benchmarks (i.e., GHERA~\cite{secureit93:online} and MSTG\&PIVAA~\cite{mstg,pivaa}) and a newly constructed CVE-based benchmark. Based on it, we gained valuable insights into these tools' performance across various dimensions, aiming to answer four research questions in \S~\ref{sec:study}. 

Our comprehensive study reveals that 
\textbf{\textit{(1)}} none of the selected SAST tools fully cover the 67 general/common vulnerability types, with the highest coverage reaching 67\%, indicating room for improvement in their detection capabilities (RQ1).
\textbf{\textit{(2)}} The results on synthetic benchmarks show that there is a significant gap between the supported vulnerability types of these SAST tools and the types injected in these synthetic benchmarks. The highest coverage rates for GHERA and MSTG\&PIVAA are 41.18\% and 50\%, respectively (RQ1).
\textbf{\textit{(3)}} The tools mainly use the method as pattern-matching for vulnerability detection, leaving a notable gap for scenario-related logical vulnerability types found in Android-specific CVEs and GHERA, like input validation vulnerabilities. (RQ2)
\textbf{\textit{(4)}} Due to the various support statuses of unified vulnerability types for these tools, their detection results cannot be quantitatively compared across different tools. Instead, we can only independently investigate the detection capability of each tool on these benchmarks.
Granularity issues in pattern matching, a lack of code context, and analysis failure are the underlying causes of the tools' effectiveness; therefore, the tools perform similarly on both synthetic and real-world benchmarks in our study (RQ3).
\textbf{\textit{(5)}} In terms of time performance, the bytecode-based SAST tools scan faster than most SAST tools that employ source code analysis (RQ4). 
Finally, we also discussed and highlighted suggestions for different stakeholders.

In summary, we made the following contributions.
\begin{itemize}
    \item To the best of our knowledge, we are the first to build a unified platform, named \tool, for evaluating SAST tools for Android, which combines the detection capability of 11 selected SAST tools by making substantial efforts to unify vulnerability types, including 67 general/common types as a taxonomy, and normalize vulnerability reports with five person-months. Additionally, \tool boasts 4,000 more lines of {Python} code.
    \item To comprehensively evaluate Android SAST tools, we constructed a new real-world benchmark based on finely filtered 292,776 CVE entries, comprising {250} Android-specific CVEs and {229} APKs, and 34 vulnerability types.
    \item Based on \tool, the existing synthetic benchmarks, and newly-constructed CVE benchmarks, we further comprehensively evaluated the detection capability of the 11 tools from different dimensions such as type coverage, type consistency, detection effectiveness, and time performance. We finally discuss some specific and useful suggestions for tool developers and users.
\end{itemize}

We have released all relevant data {and code} used in our study on GitHub~\cite{androida69:online}.

\begin{figure*}
\centering   
\includegraphics[width=0.9\textwidth]{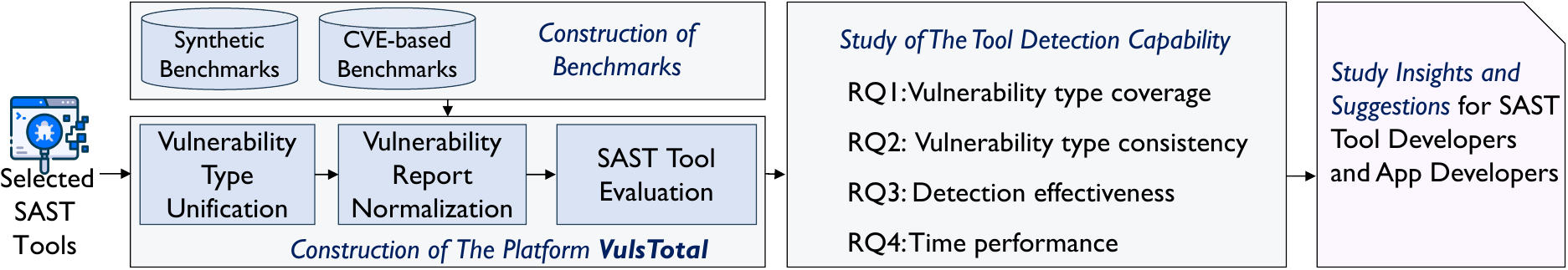}
\centering
\caption{Overview of our study.}
\label{fig: overview}
\end{figure*}

\section{Overview of Our Study}
This section introduces the key parts of our empirical study. As shown in~\Cref{fig: overview}, we first introduce the criteria for tool selection. 
Next, we describe platform construction steps involving vulnerability type unification and report normalization. 
Lastly, we discuss the details of benchmark construction.

\begin{table}[]
\caption{{Three sets of keywords used for tool collection.}}
\label{tab:keywords sets}
\centering
{\begin{tabular}{@{}lll@{}}
\toprule
\textbf{Android-specific}   & \textbf{Constraints on tools}        & \textbf{Research objectives}             \\ \midrule
APP                & Security Analysis       & Tools                        \\
Android            & Vulnerability Detection & Effectiveness Analysis         \\
Mobile Application & Static Analysis         & Systematic Literature Review \\
                   & Taint Analysis          &                              \\ \bottomrule
\end{tabular}}
\end{table}

\subsection{SAST Tool Selection}\label{sec:tool_selection}
To thoroughly evaluate the vulnerability detection capabilities of Android SAST tools, we sought out a diverse set of SAST tools from both academic and industrial domains. 
{Specifically, we scoped our research to Android SAST tools and established a dynamic and iterative process for crafting keyword sets 
which are displayed in~\Cref{tab:keywords sets}.}
{We primarily searched tools from recent literature and conducted a systematic literature review (SLR) following well-established guidelines~\cite{zhan2021research,wohlin2014guidelines,keele2007guidelines} to ensure comprehensiveness and systematicness.}
{Using the three sets of keywords from~\Cref{tab:keywords sets}, we applied logical \textit{OR} within each set and logical \textit{AND} between sets to form precise search strings.
Further, we deeply mined ACM~\cite{acmDigitalLibrary}, IEEE~\cite{ieeeDigitalLibrary}, ScienceDirect~\cite{sciencedirectScienceDirectcomScience}, SpringerLink~\cite{springerHomeSpringerLink}, and DBLP~\cite{unitrierDblpComputer} to conduct advanced search using search strings, strictly screen, and finally lock 7 core literatures~\cite{reaves2016droid,zhang2021analyzing,senanayake2023android,chen2022ausera,pauck2018android,ranganath2020free,kulkarni2018open}. The entire process was conducted by the first author, with co-authors performing cross-validation to ensure accuracy.}
{We further retrieved Android SAST tools on GitHub using the above search strings and sorted the results by star numbers. We focused on collecting tools exceeding 10 stars, ensuring the inclusion of relatively popular and widely recognized tools.}
{We conclude by obtaining a tool list from two prominent websites, including NIST~\cite{NIST} and Gartner~\cite{gartner}, using the above search strings for searching as a supplement.}
After collating data and filtering out duplicate entries, we identified 99 pertinent SAST tools in the Android vulnerability research domain, spanning both industry and academia {(all details of tool lists and screening process are available in GitHub~\cite{androida69:online}).}
To facilitate the selection and comparison of Android SAST tools for our study, we designed six selection criteria as follows:
\begin{table}[!t]
\caption{Tool profile. ``\# Stars'' indicates the number of GitHub stars. ``M.'' refers to whether the tool is maintained. ``B.\textbar S.'' denotes source code or bytecode analysis. ``Syn.\textbar Sem.'' denotes syntax-based or semantic-based core technologies.}
\centering
\label{tab: Tools Attribute}
\scalebox{0.82}{
\begin{tabular}{@{}cccccccc@{}}
\toprule
\textbf{Tool} & \textbf{\# Stars} & \textbf{Last Update} & \textbf{Version} & \textbf{Language} & \textbf{M.} & \textbf{B.\textbar S.} & {\textbf{Syn.\textbar Sem.}} \\ \midrule
\textbf{MobSF}      & 15.4k & 12/04/2023 & v3.6.0-Beta  & Python    & \Checkmark      &  S.     &  {Syn.}   \\
\textbf{QARK}       & 3.1k  & 04/05/2019 & v0.9-Alpha.1 & Python    & \XSolidBrush    &  S.     &  {Syn.}   \\
\textbf{AndroBugs}  & 1.1k  & 11/12/2015 & v1.0.0       & Python    & \XSolidBrush    &  B.     &  {Sem.}   \\
\textbf{APKHunt}    & 622   & 07/05/2023 & 07/05/2023   & Go        & \Checkmark      &  S.     &  {Syn.}   \\
\textbf{SUPER}      & 411   & 12/10/2018 & 0.5.1        & Rust      & \XSolidBrush    &  S.     &  {Syn.}   \\
\textbf{JAADAS}     & 338   & 04/12/2017 & 0.1-Alpha    & Java, Scala & \XSolidBrush  &  B.     &  {Sem.}   \\
\textbf{DroidStatx} & 115   & 12/09/2018 & 12/09/2018   & Python    & \XSolidBrush    &  B.     &  {Sem.}   \\
\textbf{Marvin}     & 68    & 11/23/2018 & 0.1-Alpha    & Python    & \XSolidBrush    &  B.     &  {Sem.}   \\
\textbf{Trueseeing} & 52    & 11/24/2023 & 2.1.9        & Python    & \Checkmark      &  B.     &  {Sem.}   \\
\textbf{AUSERA}     & 30    & 10/09/2023 & 10/09/2023   & Java, Python & \Checkmark   &  B.     &  {Sem.}   \\
\textbf{SPECK}      & 11    & 10/10/2023 & 10/23/2022   & Python    & \Checkmark      &  S.     &  {Syn.}   \\ \bottomrule
\end{tabular}}
\end{table}

\begin{figure}
\centering   
\includegraphics[width=0.45\textwidth]{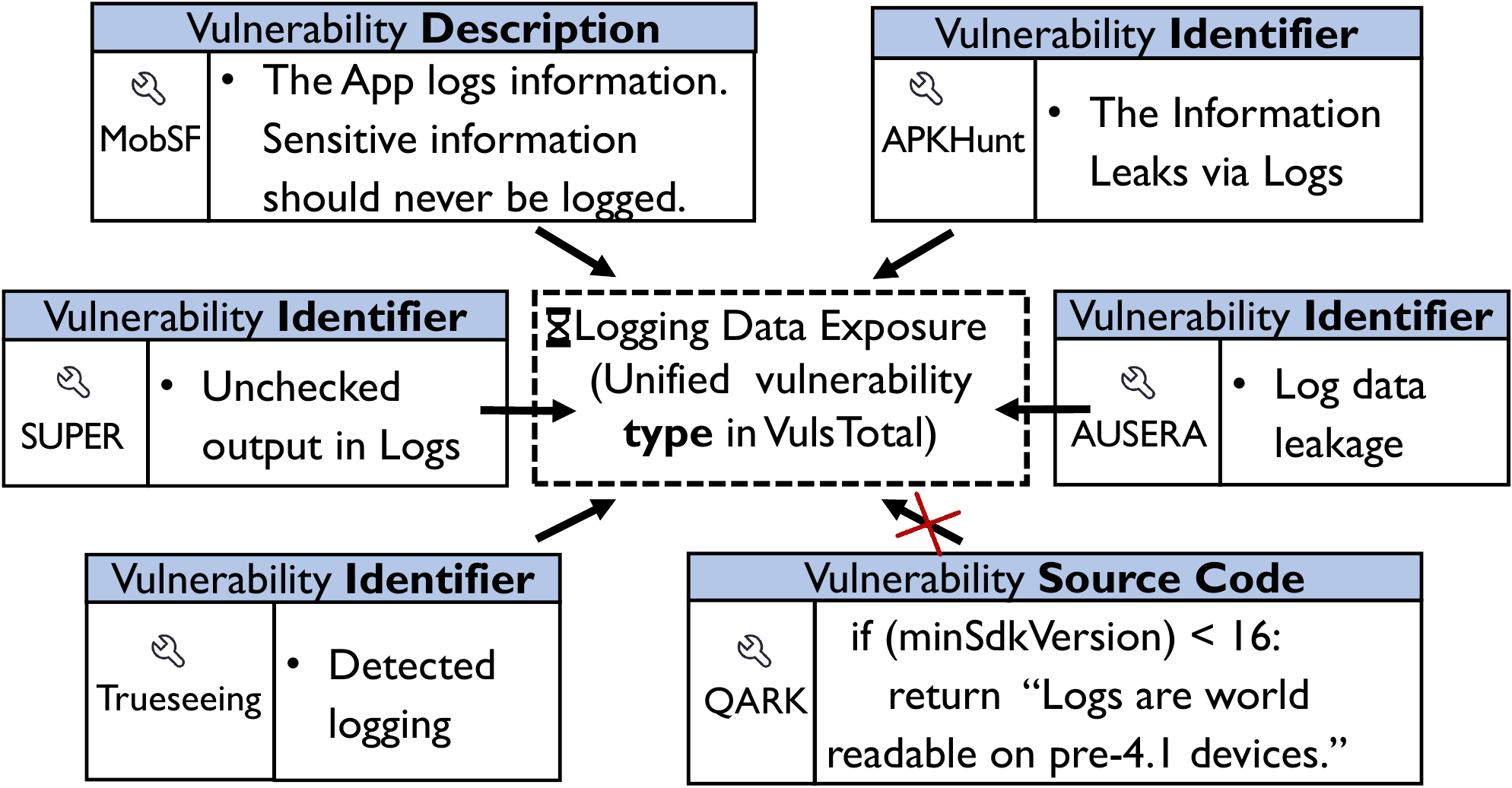}
\caption{Example of mapping unified vulnerability types.}
\label{fig:unified rules example}
\end{figure}

\noindent\textit{\textbf{\ding{172} Free of charge and transparent.}} 
The Android SAST tools must be free of charge.
While commercial tools are indeed prevalent in the industry, they often entail substantial costs, which would be prohibitive for our large-scale experiment. Additionally, since we attempted to explore the internal implementation of the tool candidates, we filtered out {47} tools that are not transparent or free, such as Quixxi~\cite{Quixxi}, ImmuniWeb~\cite{immuniweb}, and Checkmarx SAST~\cite{checkmarx}.

\noindent\textit{\textbf{\ding{173} {GitHub stars.}}} 
We tailed the star number for all tools available on GitHub and filtered out tools with fewer than 10 stars to focus on more widely recognized and potentially more established tools. We finally excluded 1 tool (i.e., WeChecker~\cite{TRUEJASO4:online}).

\noindent {{\textit{\textbf{\ding{174} Available documentation and usability.}}}} 
The Android SAST tools must be operational and accompanied by available documentation, eliminating the human bias introduced by the efforts required to discover how to build and use them. Thus, we filtered out 7 tools that lacked proper documentation or not working, such as DroidLegacy~\cite{srldroid31:online} (lack of usage docs).

\noindent {{\textit{\textbf{\ding{175} Tools compatible with APK files.}}} 
As the APK files provide a comprehensive representation of an Android application, aiding in more realistic vulnerability discovery and analysis, we filtered out 10 tools that do not support APK files as input, such as Android Check~\cite{noveogro46:online} and FindSecurityBugs~\cite{FindSecu63:online}.}

\noindent {\textit{\textbf{\ding{176} Command-line interface.}}} 
Given our objective of automating large-scale scans, while ensuring seamless integration of tool functions onto our provided unified platform \tool, we tend to choose tools that provide command-line interfaces. Web-UI-based tools without programmable API functionality are impractical. Therefore, we filtered out 2 tools. As an illustration, Aparoid~\cite{stefan2296:online} was excluded due to the lack of API integration, contrasting with tools like MobSF~\cite{MobSF} which inherently include API, both were Web-UI-based.

\noindent \textit{\textbf{\ding{177} Generalized vulnerability detection.}} 
We aim to understand the extent of coverage for various vulnerability types by current Android SAST tools. 
Thus, we focused on tools that offer comprehensive and general coverage across various vulnerability types.
Therefore, we excluded {21} tools that are designed to detect specific vulnerability types, such as SMV-Hunter~\cite{greenwood2014smv} (detecting SSL/TLS MITM vulnerabilities only), 
CogniCrypt~\cite{kruger2019crysl} (detecting vulnerable cryptographic API usage only), 
and FlowDroid~\cite{arzt2014flowdroid} ({using taint analysis to detect vulnerability types related to sensitive data}). Indeed, numerous empirical studies are dedicated to the evaluation of tools designed for the detection of specific types~\cite{zhang2021analyzing,kruger2019crysl,arzt2014flowdroid}.

Finally, we obtained 11 Android SAST tools: MobSF~\cite{MobSF}, AndroBugs~\cite{AndroBugs}, QARK~\cite{QARK}, APKHunt~\cite{APKHunt:online}, SUPER~\cite{SUPER}, JAADAS~\cite{JAADAS}, DroidStatx~\cite{droidstatx:online}, Marvin~\cite{Marvin}, Trueseeing~\cite{trueseeing:online}, AUSERA~\cite{chen2022ausera}, and SPECK~\cite{SPECK}. 
We have uploaded the full candidate SAST tool list~\cite{androida69:online} and all the detailed information. 
{\Cref{tab: Tools Attribute} provides a distilled yet holistic view of the key attributes of each tool, including the star number on GitHub, the last updated date, version, programming language, whether still maintained or not, analysis based on source code or bytecode, {and core techniques,}
thus enabling a systematic comparison and analysis of their potential effectiveness.}
{We next outline the core techniques of these tools from a structured perspective.}

{\textbf{Core techniques within selected tools.}}
These tools can be divided into two categories based on the analysis objectives: \textbf{Source code analysis} and \textbf{Bytecode analysis}. 
Upon obtaining the analysis objects, tools employ \textbf{Syntax-based} or \textbf{Semantic-based} technologies to detect vulnerabilities.
Syntax-based tools identify potential threats through predefined vulnerability patterns such as sensitive APIs using techniques including regular expression matching, string matching, and AST (Abstract Syntax Trees) matching.
Semantic-based tools usually involve control-flow and data-flow analysis to track execution paths and examine data flows. 
{Refer to~\Cref{tab: Tools Attribute},} among the selected tools, the source-code analysis tools include APKHunt, SUPER, SPECK, MobSF, and QARK. 
The first three tools use string-based pattern matching on decompiled code, whereas MobSF and QARK employ AST-based pattern matching. 
Bytecode analysis tools generally leverage existing SAST frameworks for semantic-based analysis. Examples include AndroBugs, which uses a modified version of Androguard~\cite{androguard:online}; DroidStatx, which implements customized control and data-flow analysis based on Androguard; JAADAS, which employs Soot~\cite{soot} and HEROS~\cite{heros} for taint and reachability analysis; Marvin, which integrates Soot and SAAF~\cite{saaf}; AUSERA, which combines Soot and FlowDroid~\cite{FlowDroidtools}; and TrueSeeing, which deploys proprietary data-flow analysis. 

\subsection{Construction of The Platform VulsTotal}\label{sec: platform}
Based on the two steps above, we design and implement the platform by introducing the following key phases.

\subsubsection{\textbf{Vulnerability Type Unification}}\label{sec:detection_rule_unification}
Given that SAST tools often introduce their own supported vulnerability {identifiers}, there is a notable challenge in the automated comparison among different tools. For instance, for the same type mentioned in~\Cref{fig:unified rules example}, AUSERA uses the identifier ``Logging data leakage'' for the log data exposure vulnerability, while SUPER employs ``Unchecked output in Logs''. This discrepancy poses difficulty in automatically determining whether a given SAST tool successfully identifies a specific vulnerability type. 
As such, there is a need for a unified taxonomy that can streamline the process of comparing different SAST tools. 
To address this, we conducted a two-phase manual review by engaging three co-authors for the vulnerability {identifiers} unification: 
\ding{172} {Collection of supported vulnerability identifiers:} Since none of the 11 selected tools provided well-documented identifier sets, we manually reviewed their documentation, configuration files, and source code. This review involved extracting the vulnerability identifiers/descriptions and the corresponding detection rules from each tool. Consequently, we obtained the vulnerability identifier sets from the selected tools, each includes the vulnerability identifiers, descriptions, and the corresponding source code snippets that were implemented to detect these vulnerability types.
\ding{173} {Construction of a unified taxonomy:} The second phase involved constructing a unified taxonomy using the collected vulnerability identifier sets above. 

Two key challenges arose during this phase: 
\textbf{\textit{C1: Ambiguity in vulnerability descriptions}}. Some tools use vague or non-descriptive vulnerability identifiers, making it difficult to determine the vulnerability types they support. 
As depicted in~\Cref{fig:unified rules example}, we discovered that 6 tools (such as AUSERA) can detect log data exposure with similar vulnerability identifiers and MobSF maintains a vulnerability description to present such type rather than an identifier. Contrarily, QARK lacks both, which necessitates diving into its source code to comprehend the implementation of its detection rules. 
We then examined whether the trigger code detected similar vulnerability features as \texttt{Log.(v|d|i|w|e|f|s)} {which is a regular expression and indicative of log data exposure vulnerability} to confirm the mapping results.
{
After a thorough review, we unified the identifiers from the other five tools as ``Log Data Exposure'', since QARK's Log vulnerability identified did not align with them.}
\textbf{\textit{C2: Variance in granularities}}. The granularity of the vulnerability identifiers varied among tools. For instance, for cryptographic vulnerabilities, SUPER's description was less detailed than that of other tools. SUPER merely used ``Weak Algorithm'' as a vulnerability identifier, while others used more fine-grained descriptions, such as ``AES encryption issues'' by AUSERA. To resolve this, we delved deeper into their corresponding code implementation to ascertain the vulnerability types they supported. 
After overcoming these challenges, which took us 3 person-months of rigorous type implementation review, we built the unified taxonomy by combining unified vulnerability types. \textbf{Vulnerability types are included only if supported by at least two tools.} To enhance the clarity of types, we renamed them by using a unified {identifier} to clearly reflect the root causes they represent. 
In the end, as displayed in~\Cref{tab: vulnerability taxonomy}, we established a unified taxonomy including 67 vulnerability types. 
{To increase clarity and navigability, inspired by Chen et al.'s taxonomy~\cite{chen2022ausera}, we grouped the 67 distinct types into 5 broader categories.}
{
We manually categorized each type into five categories based on specific descriptions provided by OWASP Top 10. Specifically, three co-authors independently reviewed the ``Security Weaknesses'' section of each OWASP risk page~\cite{owaspOWASPMobile} and the detection rules (source code and rule documentation). Afterward, we conducted discussions and cross-validation to ensure consistency.}
This structure enhances the identification of vulnerability types and offers an overarching view of the tools' detection capabilities. This unified taxonomy serves as a reference point for \tool, facilitating an automated comparison of the detection capabilities of the selected tools.

\subsubsection{\textbf{Vulnerability Report Normalization}}\label{sec:report_normalization}
The challenge we encountered involved disparate report formats and contents across various SAST tools, ranging from HTML web pages and terminal outputs to TXT and JSON files, impeding large-scale automated analysis. 
For instance, MobSF generates HTML reports, whereas SPECK prints reports in the terminal. Tools such as SUPER offer selectable formats including JSON and TXT. 
To overcome this, we modified the source code of these tools, aligning their report output to a consistent and processable file form. Importantly, these modifications did not affect the tools' detection logic, ensuring the authenticity of the detection results.
Consequently, this normalization process ensured a uniform reporting format {(i.e., TXT)}, facilitating a fair comparison and evaluation of various tools.

Next, we turned our attention to aligning report contents. The disparity in vulnerability types and descriptions, along with information unrelated to vulnerabilities in the reports, added another layer of complexity to our analysis. 
To solve this, we built a separate parser for each tool. These parsers were made to carefully pick out the vulnerability types and descriptions from the reports by regular matching, while also removing unnecessary information. Therefore, we achieved standardized and simplified content for all tool reports. This made it convenient to compare and understand the detection results from different tools. 

{We enhanced \tool with automation for scanning multiple APKs and integrated each tool's vulnerability detection.}
Ultimately, we choose the latest successfully configured version of each tool listed in \Cref{tab: Tools Attribute} and use their default configuration. 
After scanning with detection interfaces and using parsed results, we mapped vulnerabilities to corresponding types in taxonomy (\Cref{tab: vulnerability taxonomy}), producing a standardized result report.
{It is worth noting that vulnerability type unification relies on the vulnerability mapping database, which is extensible.}

\begin{table}
\caption{The results of 67 unified vulnerability types. (Types in the gray indicate that they were only detected by two of the tools. ``\# Out of scope'' indicates the number of security alerts rather than vulnerability types. )}
\label{tab: vulnerability taxonomy}
\centering
\scalebox{0.67}{
\begin{tabular}{c|l|c|c|c|c|c|c|c|c|c|c|c}
\hline
\textbf{Category} & \textbf{Unified Vulnerability Types} & \begin{tabular}[c]{@{}c@{}}\rotatebox{90}{\textbf{MobSF}}\end{tabular} & \begin{tabular}[c]{@{}c@{}}\rotatebox{90}{\textbf{QARK}}\end{tabular} & \begin{tabular}[c]{@{}c@{}}\rotatebox{90}{\textbf{AndroBugs}}\end{tabular} & \begin{tabular}[c]{@{}c@{}}\rotatebox{90}{\textbf{APKHunt}}\end{tabular} & \begin{tabular}[c]{@{}c@{}}\rotatebox{90}{\textbf{SUPER}}\end{tabular} & \begin{tabular}[c]{@{}c@{}}\rotatebox{90}{\textbf{JAADAS}}\end{tabular} & \begin{tabular}[c]{@{}c@{}}\rotatebox{90}{\textbf{Droidstatx}}\end{tabular} & \begin{tabular}[c]{@{}c@{}}\rotatebox{90}{\textbf{Marvin}}\end{tabular} & \begin{tabular}[c]{@{}c@{}}\rotatebox{90}{\textbf{Trueseeing}}\end{tabular} & \begin{tabular}[c]{@{}c@{}}\rotatebox{90}{\textbf{AUSERA}}\end{tabular} & \begin{tabular}[c]{@{}c@{}}\rotatebox{90}{\textbf{SPECK}}\end{tabular} \\ \hline
\multirow{14}{*}{\begin{tabular}[c]{@{}c@{}}Sensitive \\ Data \\ Exposure \\ Risks\end{tabular}} 
 & Webview Password Exposure &  &  &  &  &  & $\star$ &  & $\star$ &  & $\star$ &  \\ \cline{2-13} 
 & Logging Data Exposure & $\star$ &  &  & $\star$ & $\star$ &  &  &  & $\star$ & $\star$ &  \\ \cline{2-13} 
 & External/Internal Data Exposure & $\star$ &  & $\star$ & $\star$ & $\star$ &  &  &  &  & $\star$ & $\star$ \\ \cline{2-13} 
 & \cellcolor{gray!30} Cache Data Disclosure &  &  &  & $\star$ &  &  &  &  &  &  & $\star$ \\ \cline{2-13} 
 & Temp File Data Exposure & $\star$ &  &  & $\star$ & $\star$ &  &  &  &  & $\star$ &  \\ \cline{2-13} 
 & \cellcolor{gray!30} SQLite Data Exposure &  &  &  & $\star$ &  &  &  &  &  & $\star$ &  \\ \cline{2-13} 
 & SMS Data Exposure & $\star$ &  & $\star$ &  & $\star$ &  &  &  & $\star$ & $\star$ & $\star$ \\ \cline{2-13} 
 & \cellcolor{gray!30} Clipboard Data Exposure & $\star$ &  &  & $\star$ &  &  &  &  &  &  &  \\ \cline{2-13} 
 & Hardcoded IP Exposure & $\star$ &  &  & $\star$ & $\star$ &  &  &  & $\star$ &  &  \\ \cline{2-13} 
 & \cellcolor{gray!30} Hardcoded Email Exposure &  &  &  & $\star$ & $\star$ &  &  &  &  &  &  \\ \cline{2-13} 
 & Device ID Exposure &  &  & $\star$ &  & $\star$ &  &  &  & $\star$ &  &  \\ \cline{2-13} 
 & \cellcolor{gray!30} Android ID Exposure &  &  & $\star$ &  &  &  &  &  & $\star$ &  &  \\ \cline{2-13} 
 & Hardcoded URL Exposure &  &  &  & $\star$ & $\star$ &  &  &  & $\star$ &  &  \\ \cline{2-13} 
 & Hardcoded Sensitive Data Exposure & $\star$ &  &  & $\star$ &  &  &  & $\star$ & $\star$ &  &  \\ \hline
\multirow{11}{*}{\begin{tabular}[c]{@{}c@{}}Insufficient \\ Encryption \\ Risks\end{tabular}} 
 & Insecure  Base64 Encryption &  &  & $\star$ &  & $\star$ &  &  &  &  & $\star$ &  \\ \cline{2-13} 
 & \cellcolor{gray!30} Insecure Blowfish Encryption & $\star$ &  &  &  &  &  &  &  &  & $\star$ &  \\ \cline{2-13} 
 & Improper Handle DES Encryption & $\star$ & $\star$ &  & $\star$ & $\star$ &  & $\star$ & $\star$ & $\star$ & $\star$ & $\star$ \\ \cline{2-13} 
 & Improper Handle AES Encryption & $\star$ & $\star$ &  & $\star$ & $\star$ &  & $\star$ & $\star$ & $\star$ & $\star$ & $\star$ \\ \cline{2-13} 
 & Improper Handle RSA Encryption & $\star$ &  &  &  & $\star$ & $\star$ &  &  &  & $\star$ & $\star$ \\ \cline{2-13} 
 & Improper Handle  RC4 Encryption & $\star$ &  &  & $\star$ &  &  &  & $\star$ &  &  &  \\ \cline{2-13} 
 & Improper Handle Insecure Hash & $\star$ &  &  & $\star$ & $\star$ &  &  & $\star$ &  & $\star$ & $\star$ \\ \cline{2-13} 
 & Use Insecure Random & $\star$ &  &  & $\star$ & $\star$ &  &  & $\star$ &  & $\star$ & $\star$ \\ \cline{2-13} 
 & \cellcolor{gray!30} Weak CBC Cipher Modes & $\star$ &  &  & $\star$ &  &  &  &  &  &  &  \\ \cline{2-13} 
 & Hardcoded IV Issue & $\star$ &  &  & $\star$ &  &  &  & $\star$ &  &  &  \\ \cline{2-13} 
 & Improper Package Hardcoded & $\star$ & $\star$ & $\star$ & $\star$ & $\star$ &  & $\star$ &  &  & $\star$ &  \\ \hline
\multirow{12}{*}{\begin{tabular}[c]{@{}c@{}}{Security} \\ Misconfig \\ Risks\end{tabular}} 
 & \cellcolor{gray!30} Misuse Empty Pending Intent Issue &  & $\star$ &  &  &  &  &  &  &  & $\star$ &  \\ \cline{2-13} 
 & Improper Receiver Registration &  &  &  &  &  &  & $\star$ & $\star$ &  & $\star$ &  \\ \cline{2-13} 
 & Misuse Implicit Intent Issue &  &  &  & $\star$ &  & $\star$ &  & $\star$ &  & $\star$ & $\star$ \\ \cline{2-13} 
 & Exported Not Protected Components & $\star$ & $\star$ & $\star$ & $\star$ & $\star$ &  & $\star$ & $\star$ & $\star$ & $\star$ & $\star$ \\ \cline{2-13} 
 & Unprotected Content Provider & $\star$ & $\star$ & $\star$ & $\star$ & $\star$ &  & $\star$ & $\star$ & $\star$ & $\star$ & $\star$ \\ \cline{2-13} 
 & \cellcolor{gray!30} Sticky Broadcast Intent Issue &  & $\star$ &  &  &  &  &  & $\star$ &  &  &  \\ \cline{2-13} 
 & \cellcolor{gray!30} ContentProvider Permissions Issue & $\star$ & $\star$ &  &  &  &  &  &  &  &  &  \\ \cline{2-13}
 & Manifest Screenshot Harvest & $\star$ &  & $\star$ & $\star$ &  &  & $\star$ &  &  & $\star$ &  \\ \cline{2-13} 
 & Manifest Backup Issue & $\star$ & $\star$ & $\star$ & $\star$ & $\star$ & $\star$ & $\star$ & $\star$ & $\star$ & $\star$ &  \\ \cline{2-13} 
 & Manifest Debug Issue & $\star$ & $\star$ & $\star$ & $\star$ & $\star$ & $\star$ & $\star$ & $\star$ & $\star$ & $\star$ &  \\ \cline{2-13} 
 & Mode World Storage Readable Issue & $\star$ & $\star$ & $\star$ & $\star$ & $\star$ & $\star$ &  & $\star$ & $\star$ & $\star$ & $\star$ \\ \cline{2-13} 
 & Mode World Storage Writable Issue & $\star$ & $\star$ & $\star$ & $\star$ & $\star$ & $\star$ &  & $\star$ & $\star$ & $\star$ & $\star$  \\ \hline
\multirow{8}{*}{\begin{tabular}[c]{@{}c@{}}Insecure \\ Code \\ Execution \\ Risks\end{tabular}} 
 & Dynamic Code Loading Issue &  &  & $\star$ &  &  &  &  & $\star$ &  & $\star$ & $\star$ \\ \cline{2-13} 
 & Runtime Command Execution Issue &  &  & $\star$ & $\star$ & $\star$ &  &  & $\star$ &  & $\star$ &  \\ \cline{2-13} 
 & Rooted Device Detection & $\star$ &  & $\star$ & $\star$ & $\star$ &  &  &  & $\star$ &  &  \\ \cline{2-13} 
 & \cellcolor{gray!30} Super User Privileges & $\star$ &  &  &  & $\star$ &  &  &  &  &  &  \\ \cline{2-13} 
 & Sensitive Functionality (loadlibrary) &  &  & $\star$ &  &  &  &  &  & $\star$ & $\star$ &  \\ \cline{2-13} 
 & SQL Injection & $\star$ &  &  & $\star$ & $\star$ &  &  &  &  & $\star$ & $\star$ \\ \cline{2-13} 
 & Fragment Injection &  &  & $\star$ & $\star$ &  & $\star$ & $\star$ & $\star$ &  & $\star$ &  \\ \cline{2-13} 
 & ContentProvider Openfile &  &  &  &  &  &  & $\star$ & $\star$ &  & $\star$ &  \\ \hline
\multirow{22}{*}{\begin{tabular}[c]{@{}c@{}}Insecure \\ Network \\ Config \\ Risks\end{tabular}} 
 & Using HTTP Issue &  &  & $\star$ & $\star$ &  &  &  &  &  & $\star$ & $\star$ \\ \cline{2-13} 
 & ClearText Traffic Issue & $\star$ &  &  & $\star$ &  &  & $\star$ &  & $\star$ &  & $\star$ \\ \cline{2-13} 
 & \cellcolor{gray!30} Debug CA Configuration Issue & $\star$ &  &  &  &  &  &  &  &  &  & $\star$ \\ \cline{2-13} 
 & \cellcolor{gray!30} Use Expired Certificate &  &  &  &  & $\star$ &  &  &  &  & $\star$ &  \\ \cline{2-13} 
 & \cellcolor{gray!30} Use SHA1\_MD5 Certificate & $\star$ &  &  &  &  &  &  &  &  & $\star$ &  \\ \cline{2-13} 
 & \cellcolor{gray!30} Android Debug Certificate & $\star$ &  &  &  & $\star$ &  &  &  &  &  &  \\ \cline{2-13} 
 & Insecure  AllowUserCA & $\star$ &  &  & $\star$ &  &  & $\star$ &  & $\star$ &  &  \\ \cline{2-13} 
 & \cellcolor{gray!30} Use Insecure Socket &  &  &  &  &  &  &  &  &  & $\star$ & $\star$ \\ \cline{2-13} 
 & \cellcolor{gray!30} Use Firebase exposed & $\star$ &  &  & $\star$ &  &  &  &  &  &  &  \\ \cline{2-13} 
 & Use Insecure SSL Socket Factory &  &  & $\star$ & $\star$ & $\star$ &  &  & $\star$ &  &  &  \\ \cline{2-13} 
 & Use Invalid Hostname Verification &  & $\star$ & $\star$ & $\star$ &  & $\star$ & $\star$ & $\star$ &  & $\star$ &  \\ \cline{2-13} 
 & Use Invalid Server Verification &  & $\star$ & $\star$ & $\star$ &  & $\star$ & $\star$ & $\star$ &  & $\star$ &  \\ \cline{2-13} 
 & Use Allow All Hostname Verification & $\star$ & $\star$ & $\star$ & $\star$ & $\star$ & $\star$ & $\star$ & $\star$ &  & $\star$ & $\star$ \\ \cline{2-13} 
 & WebView Cert Validation Issue & $\star$ &  & $\star$ & $\star$ & $\star$ & $\star$ & $\star$ & $\star$ &  &  &  \\ \cline{2-13} 
 & \cellcolor{gray!30} Webview Sop Warning &  & $\star$ &  &  &  & $\star$ &  &  &  &  &  \\ \cline{2-13} 
 & Webview JavaScript Execution & $\star$ & $\star$ & $\star$ & $\star$ & $\star$ &  & $\star$ &  & $\star$ & $\star$ & $\star$ \\ \cline{2-13} 
 & Webview Java  Objects Exposure & $\star$ & $\star$ & $\star$ & $\star$ & $\star$ & $\star$ & $\star$ & $\star$ & $\star$ &  & $\star$ \\ \cline{2-13} 
 & \cellcolor{gray!30} Webview Insecure Load Plugin &  & $\star$ &  &  &  &  &  &  &  & $\star$ &  \\ \cline{2-13} 
 & Webview Local File Access &  & $\star$ & $\star$ & $\star$ &  & $\star$ & $\star$ & $\star$ &  & $\star$ &  \\ \cline{2-13} 
 & \cellcolor{gray!30} WebView Local File Cleanup &  &  &  & $\star$ &  &  &  &  &  &  & $\star$ \\ \cline{2-13} 
 & WebView Insecure  URL Loading &  & $\star$ &  & $\star$ &  &  & $\star$ &  &  &  &  \\ \cline{2-13} 
 & \cellcolor{gray!30} WebView Remote Debugging & $\star$ &  &  & $\star$ &  &  &  &  &  &  &  \\ \hline

\hline
\multirow{3}{*}{\textbf{Metadata}}
& \textbf{\# Overlapped vulnerability types} & 39 & 21 & 27 & 45 & 32 & 15 & 21 & 28 & 21 & 40 & 23 \\ \cline{2-13} 
& \textbf{\# Unique vulnerability types} & 12 & 2 & 5 & 15 & 0 & 3 & 4 & 3 & 5 & 1 & 4  \\ \cline{2-13} 
& {\textbf{\# Out of scope}} & 26 & 2 & 21 & 16 & 14 & 1 & 13 & 15 & 8 & 0 & 6  \\ \hline

\end{tabular}}
\end{table}

\subsection{{Construction of Benchmarks }}\label{subsec:benckmarks}
To evaluate the performance of tools, we collected two kinds of benchmarks: synthetic benchmarks (with injected vulnerabilities) and CVE-based benchmarks (with real-world vulnerabilities).

\subsubsection{\textbf{Synthetic Benchmarks}}\label{subsec:syn}
We collected synthetic benchmarks from both the academy and industry. Although many different kinds of synthetic benchmarks such as DroidBench~\cite{fritz2013highly}, ICC-Bench~\cite{wei2018amandroid}, and UBCBench~\cite{zhang2021analyzing} are widely used in the academy, they are used to evaluate the effectiveness of static taint analysis tools and not suitable under our evaluation scenario because selected SAST tools focus on detecting common vulnerability types instead of specific types. 
By referring to the evaluation and comparison results in~\cite{mitra2019benchpress}, we choose GHERA~\cite{mitra2017ghera} 
since it is a representative benchmark~\cite{mitra2017ghera} maintaining more vulnerability types and providing both benign and secure versions for each vulnerability type.

Additionally, some industry companies and institutions developed vulnerable apps by manually injecting vulnerabilities. From this side, we take MSTG app~\cite{mstg} and PIVAA app~\cite{pivaa} into account, where MSTG is maintained by OWASP and the latter one is developed by an industry company named High-Tech Bridge~\cite{HighTech62:online}.
The MSTG serves as a comprehensive resource for mobile app security testing, providing valuable insights into identifying and addressing potential vulnerabilities. Meanwhile, the {PIVAA} app showcases real-world security issues and serves as an educational tool to enhance app developers' understanding of secure coding practices.

\subsubsection{\textbf{CVE-based Benchmark}}\label{subsec:cve}
To create a verified real-world benchmark containing vulnerabilities caused by Android app developers, we chose the CVE database~\cite{CVECVE56:online} as the source, which maintains an open list of known real-world vulnerabilities found in specific software products.
\ding{172} Initially, we filtered the CVE database for entries containing the keyword ``Android'' as of 2023-09-12, which yielded 8,451 vulnerabilities.
\ding{173} In the remaining CVE entries, we found that some vulnerabilities lie in C/C++ files which are beyond our research scope. We thus filtered out 2,042 such CVE entries. 
\ding{174} To maintain the benchmark's focus on Android applications, we excluded vulnerabilities tied to multiple platforms (e.g., Windows), Android underlying components (e.g., Android media framework), generic Android tools (e.g., Jadx~\cite{jadx}) and so on. Consequently, we filtered out 4,029 CVE entries.
\ding{175} Since we focus on developer-related issues within Android apps, we also excluded 277 CVEs only related to the browser kernels, as well as those marked as controversial, disputed, or unspecified, such as CVE-2021-43512~\cite{CVECVE2048:online}.
\ding{176} After excluding cross-language vulnerabilities (such as out-of-bound errors) that do not arise from Android development defects, 2,079 Android-specific CVE entries remained.  
\ding{177} Finally, as we aim to gather as many vulnerabilities as possible about the specified app and version, we filtered out 46 CVE entries that did not specify the version information, while 2,033 CVE entries remained.

To construct a comprehensive benchmark aligned with the supported vulnerabilities across 11 tools, covering diverse vulnerability types for a thorough evaluation of Android SAST tools, 
we further refined 2,033 entries. Based on the taxonomy and the unique vulnerability types supported by each tool, we labeled the corresponding vulnerability types for these CVEs based on their descriptions and supplementary information. 
To avoid potential bias in the labeling process, detailed information on each CVE was rigorously reviewed and independently labeled by three co-authors. In case of disagreement, the final decision was made by majority voting.
{In total, we assigned 2,050 labels to 2,033 CVEs since certain CVEs include up to two types. Of these, 1,722 labels correspond to types supported by selected tools, while 328 labels correspond to types not supported by any selected tool, thus outside the study's scope.} 

Regarding the {1,722 labels}\footnote{In this paper, the term ``label'' denotes ``vulnerability instance''.} in our research scope, we attempted to download all available vulnerable APKs as described in their respective CVE entries. 
{
Specifically, we have spent substantial time and effort searching in APKPure~\cite{apkpure}, APKMonk~\cite{apkmonk}, Google Play~\cite{googleplay}, and other app markets, as well as the AndroZoo database~\cite{allix2016AndroZoo}. Finally, we found available APKs corresponding to 1,316 instances. {It is noteworthy that there is a long-tail distribution~\cite{Longtail81:online} in vulnerability types, identifying 1,143 instances just involving 3 specified vulnerability types.\footnote{The types are respectively ``Use Invalid Server Verification'', ``Use Invalid Hostname Verification'', ``Use Allow All Hostname Verification''.}}
Moreover, we could not feasibly scan all instances due to resource constraints. 
Focusing on the remaining 173 instances, we noted a maximum of 30 instances per single type.
Therefore, we opted to randomly select 30 instances from each of these three types to be included in the CVE benchmark for effectiveness evaluation. 
{Subsequently, for these three types, we incrementally added 10 instances to each type until they all reached 60, covering four different states. We continuously calculated the Recall value for all tools in these four states on the CVE benchmark and observed that across the four states, the sample variance of effectiveness\footnote{We quantified it by using B\_Recall, which is defined in~\Cref{eq:b_recall}.} for each tool on the CVE benchmark was under 0.1\%.} Based on this finding, we deduced that including all 1,143 instances versus including 90 samples (30 per type) would have a negligible effect on the final results.}

{Therefore, we chose 30 samples per type, resulting in the CVE benchmark including \textbf{250} CVEs encompassing \textbf{229} APKs, \textbf{262} vulnerability instances that covered \textbf{34} vulnerability types named \textbf{CVE-based benchmark}. All labeling data, detailed description of APK collection, Recall of four calculations, and the CVE-based benchmark are released on GitHub~\cite{androida69:online} and Zenodo~\cite{TheRealW98:online}. 
}

\section{Study of The Tool Detection Capability}\label{sec:study}
With the aid of the ability of \tool, our study addresses the following research questions to evaluate the detection capability of the 11 selected tools comprehensively.

\begin{itemize}
    \item \textbf{RQ1: (Vulnerability type coverage)} Are these SAST tools capable of covering the unified vulnerability types that are supported by \tool? What is the coverage of vulnerability types in the used benchmarks?  
    \item \textbf{RQ2: (Vulnerability type consistency)} Is the Android vulnerability landscape documented in CVE consistent with the coverage provided by the selected SAST tools? What about GHERA and MSTG\&PIVAA?
    \item \textbf{RQ3: (Detection effectiveness)} How effective are these SAST tools for vulnerability detection on different benchmarks? {How do these tools perform in terms of the same vulnerability types?}
    \item \textbf{RQ4: (Time performance)} What are the different statuses of time performance for these SAST tools? 
\end{itemize}

\begin{figure*}[]
	\centering
	\subfloat[ GHERA benchmark.]{\includegraphics[width=.32\linewidth]{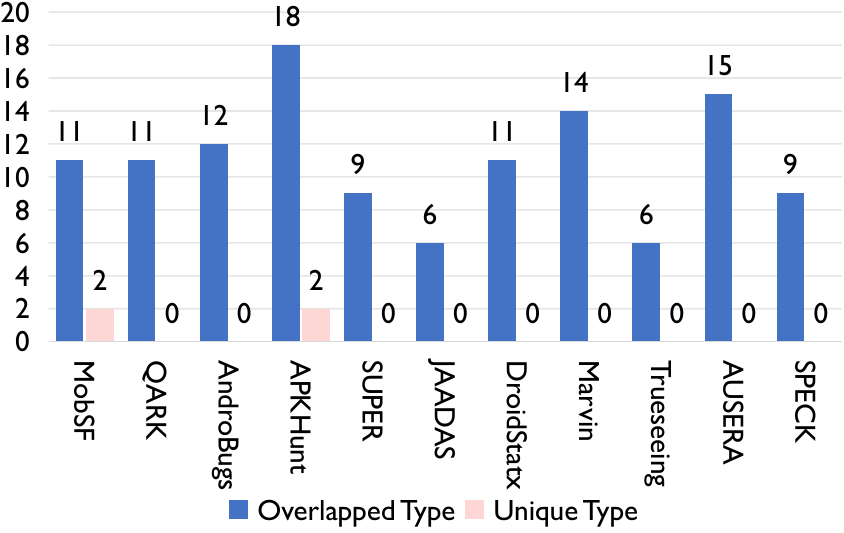}\label{fig:GHERA_cover}}\hspace{1pt}
	\subfloat[ MSTG\&PIVAA benchmark.]{\includegraphics[width=.32\linewidth]{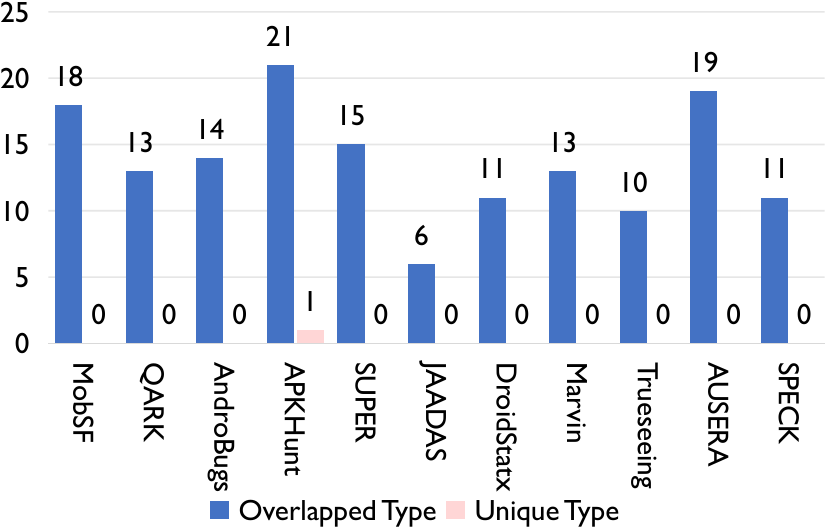}\label{fig:MSTG_cover}}\hspace{1pt}
	\subfloat[ CVE-based benchmark.]{\includegraphics[width=.32\linewidth]{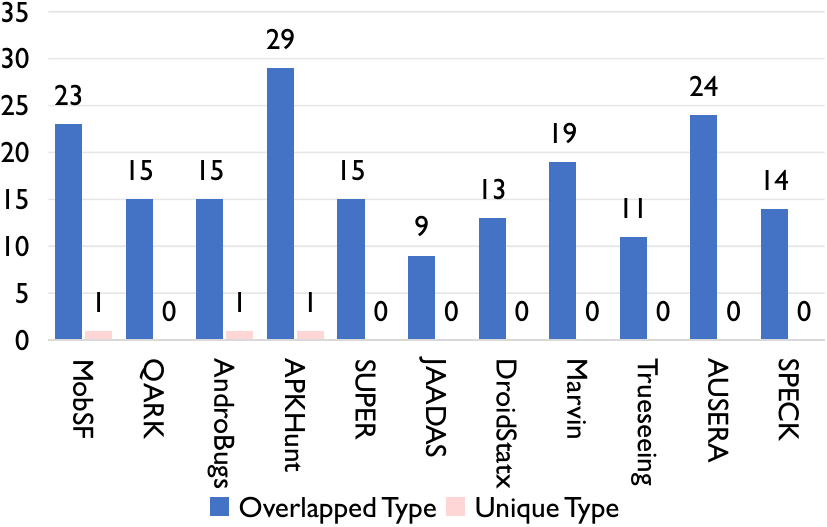}\label{fig:CVE_cover}}
    \caption{Vulnerability type coverage of Android SAST tools in different benchmarks.}
    \label{fig:Tools coverage}
\end{figure*}

\subsection{RQ1: Vulnerability type coverage}\label{sec:rq1}
\subsubsection{\textbf{Setup}}
The range of vulnerability types a SAST tool can detect serves as a significant measure of its {overall performance}. 
To this end, we aim to explore the vulnerability type coverage of each tool on the unified taxonomy and the three benchmarks collected in~\S~\ref{subsec:benckmarks}.
To achieve it, we further categorize the vulnerability {types} based on the proposed taxonomy into three groups: \ding{172} \textbf{Overlapped types:} vulnerability types {supported by multi-tools}, \ding{173} \textbf{Unique types:} vulnerability types supported only by a single tool, and \ding{174} \textbf{Unsupported types:} unsupported vulnerability types by all tools. 

\subsubsection{\textbf{Result}}

\begin{figure}
\centering   
\includegraphics[width=0.48\textwidth]{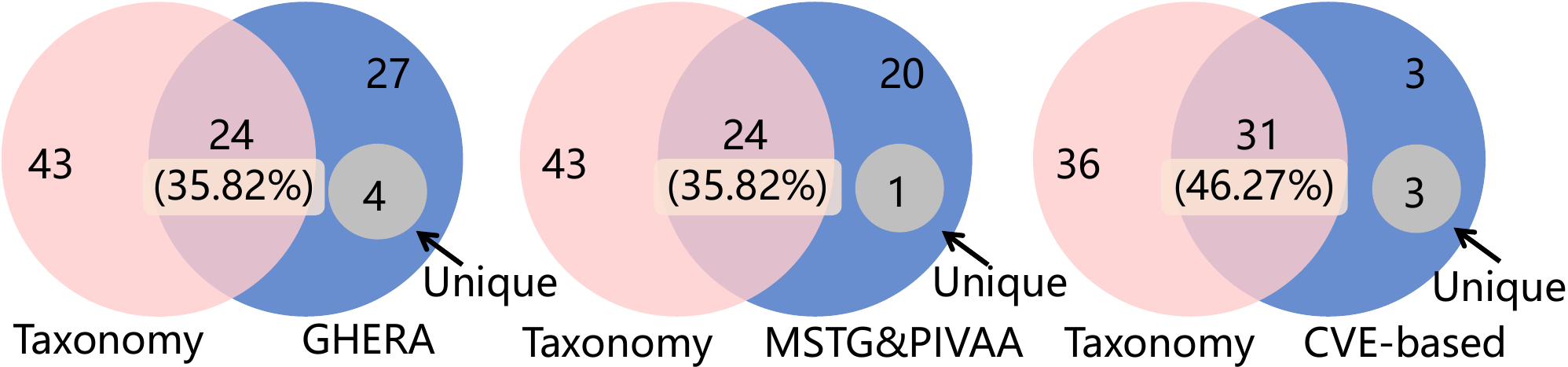}
\centering
\caption{{Vulnerability types supported by \tool (in pink) and each benchmark (in blue). ``Unique'': supported by one tool only.}}
\label{fig:Benchmark consistency}
\end{figure}

As depicted in Table~\ref{tab: vulnerability taxonomy}, we included the number of both \textbf{overlapped} and \textbf{unique} vulnerability types supported by each tool. 
{Further, types related to code-quality issues only rather than vulnerabilities, and thus out of our research scope, are also tracked and represented as ``\# Out of scope''. 
For example, the ``MANIFEST\_GCM'' supported by AndroBugs indicates that if the app's ``minSdkVersion'' is less than 9, then the app cannot use Google Cloud Messaging (GCM). This is a compatibility issue~\cite{DBLP:conf/icse/YangCFXHH23} rather than a security vulnerability, so it falls outside our research scope. }
Based on it, we found that these tools typically lack comprehensive coverage of the {overlapped} vulnerability types in \tool. Indeed, coverage tends to vary significantly among different tools.
Notably, APKHunt boasts the highest coverage at 67\% (45/67), with AUSERA coming in second at coverage of 60\% (40/67), whereas JAADAS lags with the lowest coverage, only 22\% (15/67).
{Refer to~\Cref{tab: Tools Attribute}, newer tools like APKHunt typically exhibit better coverage than older tools like JAADAS. 
This may be because newer tools can cover the vulnerability types that are constantly being newly discovered. 
However, we emphasize that the relationship between tool age and coverage does not vary linearly.}
Moreover, we discovered that there are certain types that most tools fail to support (only detected by two selected tools), which are highlighted in~\Cref{tab: vulnerability taxonomy}. 
For instance, within the ``Insecure Network Config Risks'' category, no more than three tools support 45\% (10/22) of the types. {This emphasizes the need for tool developers to focus on detecting these frequently overlooked types to enhance the comprehensiveness of their type coverage.}
Furthermore, we noticed a significant overlap in types supported by different tools, suggesting a shared understanding among developers about the significance and universality of certain types.
Specifically, nearly all tools support detecting vulnerability types in \textit{AndroidManifest.xml}, like ``Manifest Backup/Debug Issue'' (excluding SPECK) and ``Exported Not Protected Components'' (excluding JAADAS).

As displayed in~\Cref{fig:Benchmark consistency}, {the alignment between the vulnerability types injected in the existing three benchmarks and the overlapped types in taxonomy} is not as high as anticipated. The consistency percentages against GHERA, MSTG\&PIVAA, and CVE-based are 35.82\%, 35.82\%, and 46.27\%, respectively.
Further, we recorded the number of overlapped and unique types covered by each tool across the three benchmarks (refer to Figure~\ref{fig:Tools coverage}). Remarkably, no tool manages to cover all vulnerability types in three benchmarks. 
We further found that APKHunt achieves the highest coverage on all three benchmarks meanwhile, with 39\% on GHERA, 50\% on MSTG\&PIVAA, and 88\% on CVE-based respectively.
Further, AUSERA also attained the second-highest coverage across three benchmarks, with 29\% on GHERA, 43\% on MSTG\&PIVAA, and 71\% on CVE-based. However, JAADAS simultaneously achieved the lowest coverage across three benchmarks, with 12\% on GHERA, 14\% on MSTG\&PIVAA, and 26\% on CVE-based.
{Similarly, this demonstrates the suboptimal coverage of the selected tools across the three benchmarks.}

Additionally, through Figure~\ref{fig:Benchmark inconsistency}, there are still many unsupported types by all tools (45.10\%, 23/51 on GHERA and 43.18\%, 19/44, on MSTG\&PIVAA, {the detailed list is also available on GitHub~\cite{androida69:online}}). The results on synthetic benchmarks show that there is a significant gap between the supported types of these SAST tools and the types injected in these benchmarks.
This inconsistency highlights the need for a reliable benchmark {to align the coverage of types supported by existing Android SAST tools.}
{However, the vulnerability types in the CVE-based benchmark are a subset of the types supported by these SAST tools, as the latter was the baseline for the former's construction. This means that all types in CVE-based are supported by tools, so the corresponding number in Figure~\ref{fig:Benchmark inconsistency} is 0.}

\smallskip
\noindent\fbox{
\parbox{0.95\linewidth}{
\textbf{Answer to RQ1:} 
\ding{172} All evaluated tools exhibit significant gaps in their support for 67 unified vulnerability types. No single tool offers comprehensive support; the highest and lowest coverage rates are 67\% and 22\%, respectively. This highlights the imperative for a comprehensive vulnerability scan of an Android app, necessitating the collaborative use of multiple SAST tools. 
\ding{173} A disparity exists between the vulnerability types supported by these tools and those present in two synthetic benchmarks, with an inconsistency rate of 45.1\%(GHERA) and 43.18\%(MSTG\&PIVAA).}}

\subsection{RQ2: Vulnerability type consistency}\label{sec:rq2}
\subsubsection{\textbf{Setup}} 
{Firstly}, since Android-related CVEs provide insights into real-world vulnerabilities, we try to explore the consistency between the vulnerability types included in Android-specific CVEs and those supported by all 11 tools.
Based on the {final-filtered 2,050} labels from \S~\ref{subsec:cve}, we incorporated them and {46 CVEs without specified app version (involving 47 labels) excluded in \S~\ref{subsec:cve} into discussion regarding their reflection of vulnerability type distribution in CVEs.}
{Based on their corresponded vulnerability types,} we categorized the labels into two groups: \textit{1)} \textbf{\textit{Supported types}}, included in the set of vulnerability types supported by the 11 tools, and \textit{2)} \textbf{\textit{Unsupported types}}.
Moreover, refer to \S~\ref{sec:rq1}, we discovered a huge gap between the supported types and those available in the synthetic benchmarks.
We further analyzed this inconsistency here. For types injected in synthetic benchmarks, we also categorized them into two groups above.
{Finally, since the OWASP Mobile Top 10~\cite{owaspOWASPMobile} (OWASP in short) represents the top 10 prominent security risks in mobile applications, we further analyzed the consistency between the tool-supported types and those outlined in OWASP. Subsequently, we will analyze type consistency from these three perspectives.}

\begin{figure}
\centering   
\includegraphics[width=0.45\textwidth]{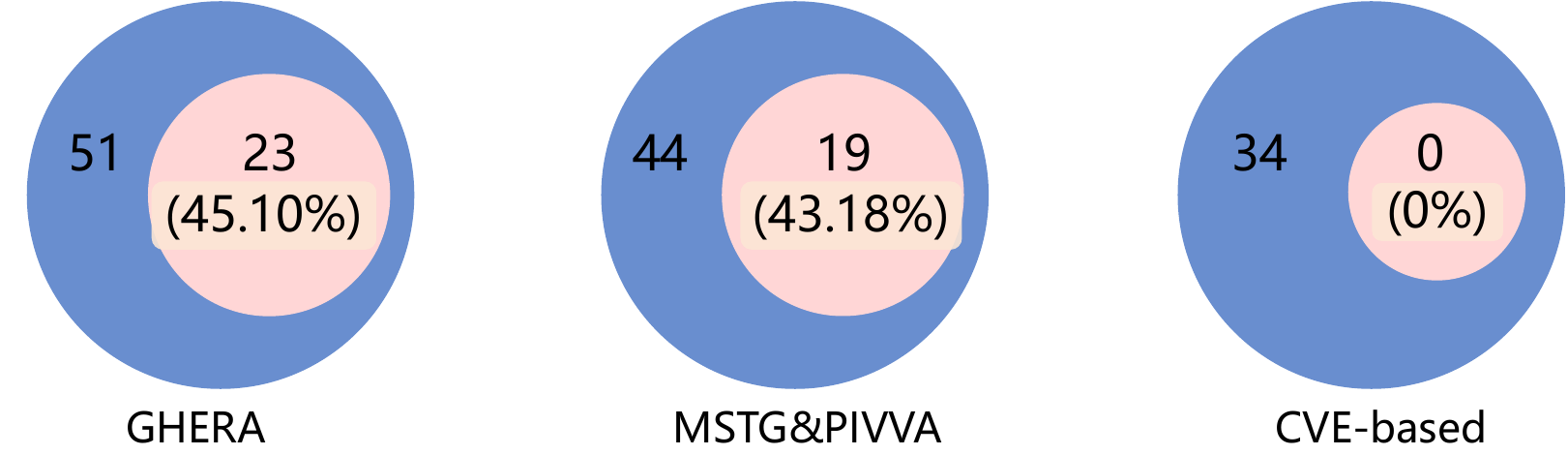}
\centering
\caption{Vulnerability types injected in each benchmark (in blue) while unsupported by all the 11 tools (in pink). }
\label{fig:Benchmark inconsistency}
\end{figure}

\subsubsection{\textbf{Result}} 
\begin{figure*}
\centering   
\includegraphics[width=0.95\textwidth]{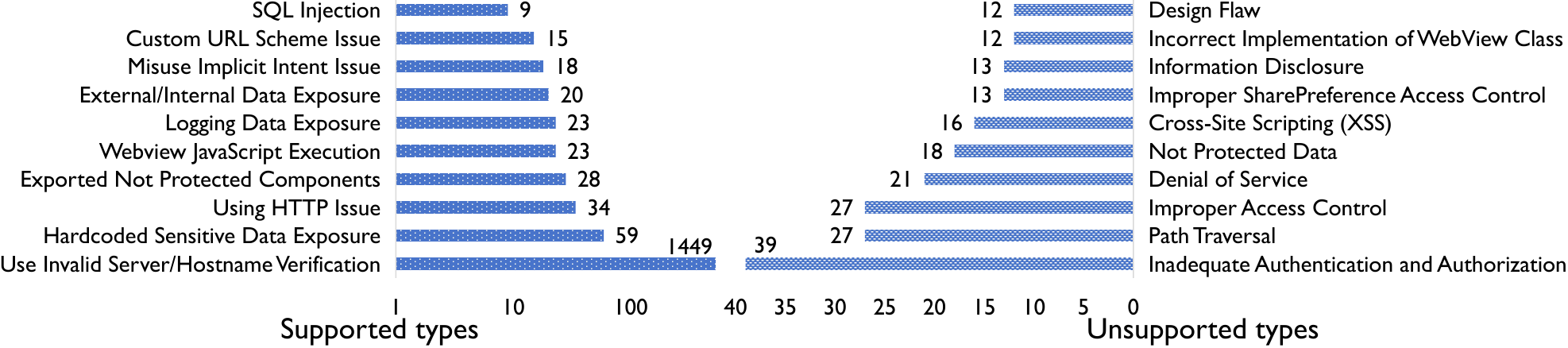}
\centering
\caption{Top 10 vulnerability types in two categories {from labeled CVEs}. Notably, there's a big difference in the number of {labels identified in CVE} from different vulnerability types in the supported types. To make it easier to understand, we used a logarithmic scale with a base of 10, increasing in multiples of 10.}
\label{fig: cve-types}
\end{figure*}

{\textbf{Android-specified CVE vulnerability type consistency.}}
We counted the {number of vulnerability types, labels, and the corresponding CVEs} in these two categories and listed them in~\Cref{tab: cve_types_num}. There are 36 supported types with 1,741 {labels} and 34 unsupported types with 356 {labels}. 
To simplify the presentation, we ranked the types in descending order based on {the number of labels in each type}. 
\Cref{fig: cve-types} shows the top 10 types and their label counts for both categories. 
We observed that among the supported types, ``Use Invalid Server/Hostname Verification''\footnote{This category contains three vulnerability types which are ``Use Invalid Server Verification'', ``Use Invalid Hostname Verification'' and ``Use Allow All Hostname Verification'' respectively.} has the highest label number at 1,449, significantly surpassing other types. The second most prevalent type is ``Hardcoded Sensitive Data Exposure'', totaling 59 labels.
Further, among the unsupported types, the most frequent is ``Inadequate Authentication and Authorization'' totaling 39, followed by ``Path Traversal'' with 27 labels.

We further classified these types based on whether they could be detected using {syntax-based or semantic-based analysis mentioned in \S~\ref{sec:tool_selection}} and found that 79\% (27/34) of the unsupported types were challenging to detect without a deep understanding of the application's scenario logic. In other words, these vulnerability types do exceed the ability of SAST tools to abstract and track complex vulnerability patterns to a certain extent.
For example, ``Lack of Input Validation'' requires tracking control and data flow to identify deficiencies in validation branching decisions, as well as conducting complex checks such as length validation and type checking according to the specific application contexts {to verify that the code is robust, well-placed, and triggered correctly in application contexts.}
Conversely, the selected tools consistently demonstrated the ability to identify vulnerability types without requiring deeper contextual understanding, such as detecting insecure encryption algorithms like ``AES/ECB'' mode, by recognizing known insecure sensitive API usage patterns.

\begin{table}\footnotesize
\caption{The \# of the vulnerability types, labels, and CVEs in two categories.}
\centering
\label{tab: cve_types_num}
\scalebox{0.9}{
\begin{tabular}{@{}lccc@{}}
\toprule
                  & \# Vulnerability Type  & \# Vulnerability Label & \# CVE \\ \midrule
Supported Types   & 36                    & 1,741                   & 1,723                  \\
Unsupported Types & 34                    & 356                     & 356                    \\ \bottomrule
\end{tabular}}
\end{table}

\begin{table}[]
\caption{{Mapping of OWASP Mobile Top 10 2024 to categories in unified taxonomy.}}
\label{tab: mapping with OWASP}
\centering
\renewcommand\arraystretch{1.3}
{\begin{tabular}{l|l}
\hline
\textbf{OWASP Mobile Top 10 2024}              & \textbf{Categories in unified taxonomy}        \\ \hline
M1: Improper Credential Usage                  & \multirow{3}{*}{Sensitive Data Exposure Risks} \\ \cline{1-1}
M6: Inadequate Privacy Controls                &                                                \\ \cline{1-1}
M9: Insecure Data Storage                      &                                                \\ \hline
M5: Insecure Communication                     & Insecure Network Config Risk                   \\ \hline
M10: Insufficient Cryptography                 & Insufficient Encryption Risks                  \\ \hline
M4: Insufficient Input/Output Validation       & \multirow{2}{*}{Insecure Code Execution Risks} \\ \cline{1-1}
\multirow{2}{*}{M8: Security Misconfiguration} &                                                \\ \cline{2-2} 
                                               & Security Misconfig Risks                       \\ \hline
M7: Insufficient Binary Protections            & None                                           \\ \hline
M2: Inadequate Supply Chain Security           & None                                           \\ \hline
M3: Insecure Authentication/Authorization      & None                                           \\ \hline
\end{tabular}}
\end{table}

{\textbf{Synthetic benchmarks vulnerability type consistency.}} 
Similarly, among the 23 vulnerability types unsupported by all 11 tools in GHERA from~\Cref{fig:Benchmark inconsistency}, there are 65\% (15/23) types that posed challenges for detection using the common methods employed by {the selected} SAST tools, which include both {syntax-based or semantic-based analysis.} 
Notably, vulnerability types injected in MSTG\&PIVAA are detectable just by {syntax-based pattern matching}.
These observations underscore the limitations of existing SAST tools that depend solely on identifying sensitive API usage through {regular-expression-based or string-based pattern matching}. It underscores the necessity for more precise pattern extracting tailored to specific types, such as privilege escalation, along with the deeper adoption of advanced detection techniques such as data and control flow analysis. These enhancements are particularly crucial for accurately identifying complex vulnerabilities that involve scenario-specific logic.
\jy{}

{\textbf{OWASP Mobile Top 10 vulnerability type consistency.}}
{Given the OWASP encompassing specific vulnerability types, we adopted its ten overarching categories as our baseline for comparison. 
Refer to~\Cref{tab: vulnerability taxonomy}, our taxonomy consolidates 67 unified types into five major categories, and we mapped these categories to OWASP categories, discovering that each category can be mapped to certain categories in OWASP, 
{indicating the taxonomy's practical significance with the high consistency with OWASP.}
Specifically, as displayed in~\Cref{tab: mapping with OWASP}, excluding M2, M3, and M7, all other OWASP-defined risks can match the categories in the taxonomy. Similarly, each unique type supported by a single tool also falls into one of the OWASP categories.
Notably, the lack of support for M3 coincides with the unsupported types {(i.e., ``Inadequate Authentication and Authorization'')} in CVEs, as shown in~\Cref{fig: cve-types}. 
This not only highlights the consistency of the distribution of vulnerability types in the real world but also emphasizes that these unsupported types should be a key focus for future tool development and optimization given their prevalence. 
The detailed mapping of types {(including those within the taxonomy and those uniquely supported by tools)} to their corresponding OWASP category is available on our GitHub~\cite{androida69:online}.
}

\noindent\fbox{
\parbox{0.95\linewidth}{
\textbf{Answer to RQ2:} {
\ding{172} {We found significant gaps between the vulnerability types included in Android-specific CVEs, those in synthetic benchmarks, and the types supported by the tools. Specifically, none of the selected tools support 34 types in Android-specific CVEs, 19 types in MSTG\&PIVAA, and 23 types in GHERA}
\ding{173} Further analysis highlights that the unsupported types are primarily those challenging for most SAST tools to cover. Specifically, 79\% of 34 unsupported types in Android-specific CVEs and 65\% of 23 unsupported vulnerability types in GHERA could not be detected using pattern matching only.
}}}

\subsection{RQ3: Detection effectiveness} \label{sec: rq3}

\subsubsection{\textbf{Setup}}
The CVE-based benchmark exhibits an uneven distribution of vulnerability instances. For instance, there are {24} instances under ``Using HTTP Issue'' but only one instance under ``Weak CBC Cipher Mode''. To ensure a comprehensive evaluation of the selected tools, we further constructed another uniform benchmark named \textbf{CVE-U} by applying an under-sampling technique~\cite{WhatIsUn32:online} to achieve a more balanced distribution of vulnerability instances. 
Specifically, we sampled a maximum of three instances for each type of vulnerability. This threshold was chosen to balance the distribution, considering the prevalent types and the limited availability of application resources.
Our analysis then compares the effectiveness of the tools using both the original, imbalanced \textbf{CVE-based} benchmark and the newly balanced \textbf{CVE-U} benchmark.

\begin{figure}
\centering   
\includegraphics[width=0.9\columnwidth]{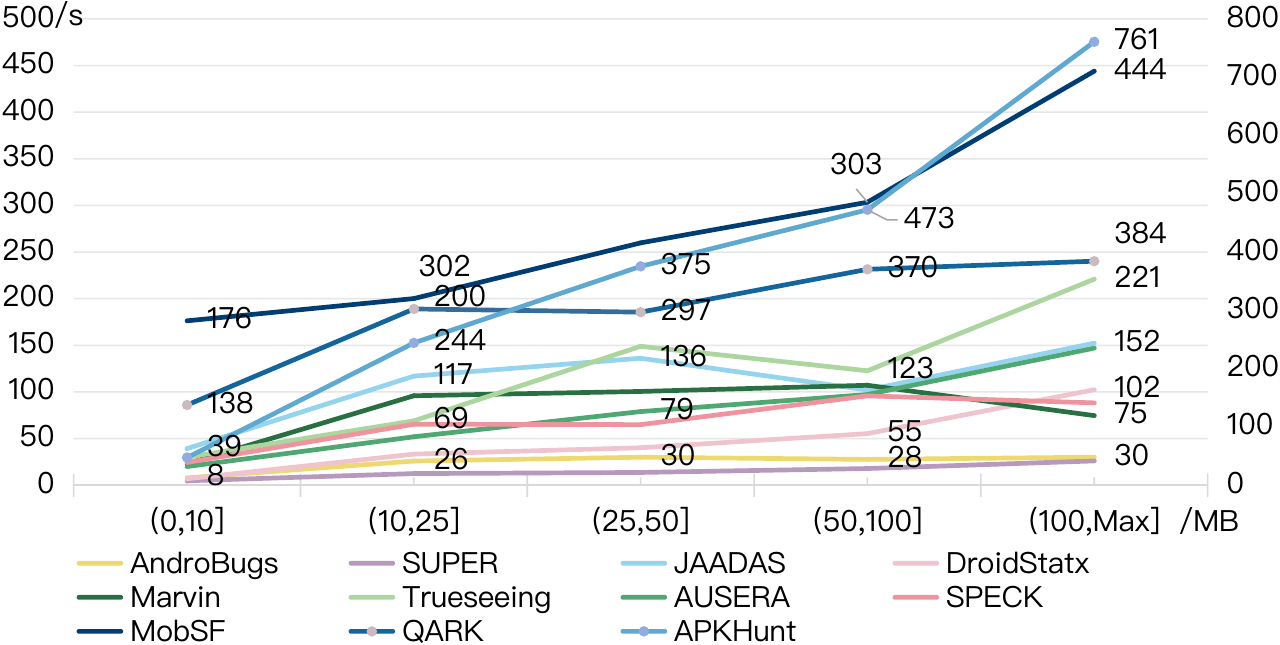}
\centering
\caption{The scanning time of each tool across different APK size intervals.}
\label{fig: time_size}
\end{figure}

To investigate the effectiveness of selected tools for vulnerability detection on these 4 different benchmarks, we leverage the platform \tool to analyze all the instances.
{Given the overlap between the tools in~\cite{chen2020empirical} and our study, we set a 15-minute timeout pre scan based on its time performance finding. Our results in~\Cref{fig: time_size} showed that all tools had a maximum average scan time below 15 minutes, confirming the validity of this timeout.}

All experiments are performed on an 8-core Linux machine with 32 GB RAM (used consistently throughout this study.)
We will discuss the effectiveness of 11 tools based on Precision, Recall, {False Positive Rate (FPR)}, and F1-score. 
\textit{\textbf{Given that ground truth is only available for known vulnerabilities in benchmarks i.e., CVE-based, CVE-U, and MSTG\&PIVAA, it is important to acknowledge that these sources cannot guarantee the absence of other vulnerabilities.}}
{Therefore, following the common practices~\cite{fse23,issta22-c}, we only calculate the customizable Recall named B\_Recall (Benchmark Recall) for them to reflect whether the selected tools could find known and documented vulnerabilities.} 
The calculation method of B\_Recall is as follows.

{
\begin{equation}
    \text{B\_Recall} = \frac{\text{\# Correctly Identified Vulns}}{\text{\# All Known Vulns in the Benchmark}} \label{eq:b_recall}
\end{equation}
}

To deeply understand the selected tools' effectiveness in {unified vulnerability types}, we further explore their detection capabilities on specific types. As detailed in~\Cref{fig:single_effectiveness}, we aggregated instances from all benchmarks for each vulnerability type, focusing on those with at least five instances to ensure a meaningful evaluation. This approach allowed us to assess the B\_Recall of tools in detecting these selected types, providing a granular view of individual tool performance and maintaining credibility by avoiding types with fewer instances.

\begin{figure*}
	\centering
	\subfloat[GHERA.]{\includegraphics[width=.24\linewidth]
    {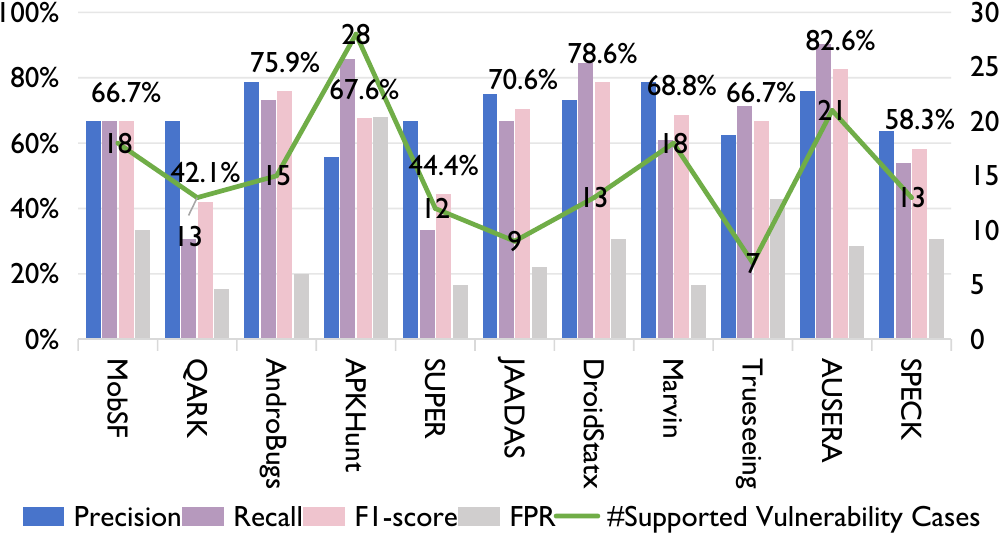}\label{fig:GHERA_effect}}\hspace{1pt}
	\subfloat[MSTG\&PIVAA.]{\includegraphics[width=.23\linewidth]{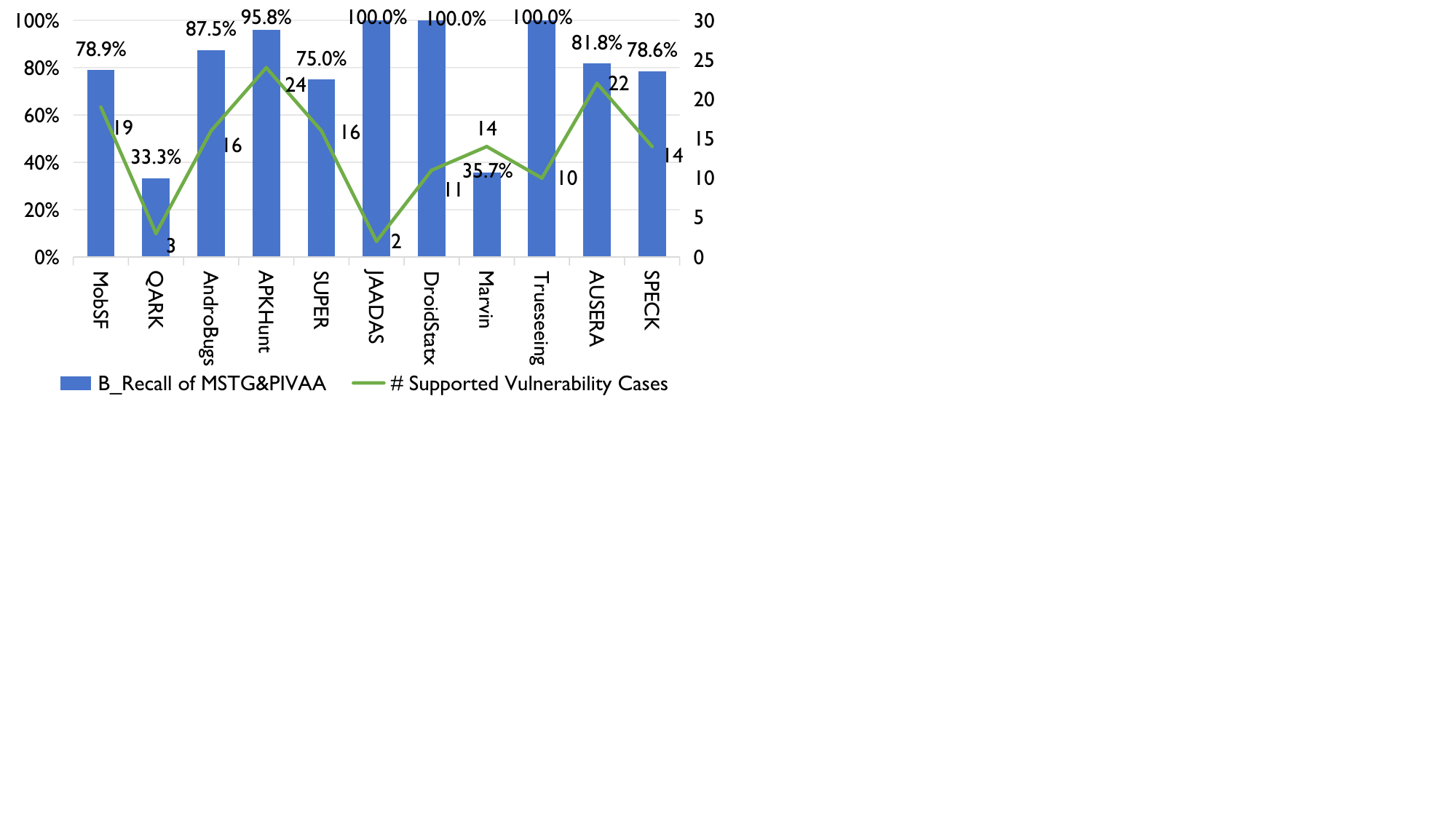}\label{fig:MSTG_effect}}\hspace{1pt}
	\subfloat[CVE-based.]{\includegraphics[width=.23\linewidth]{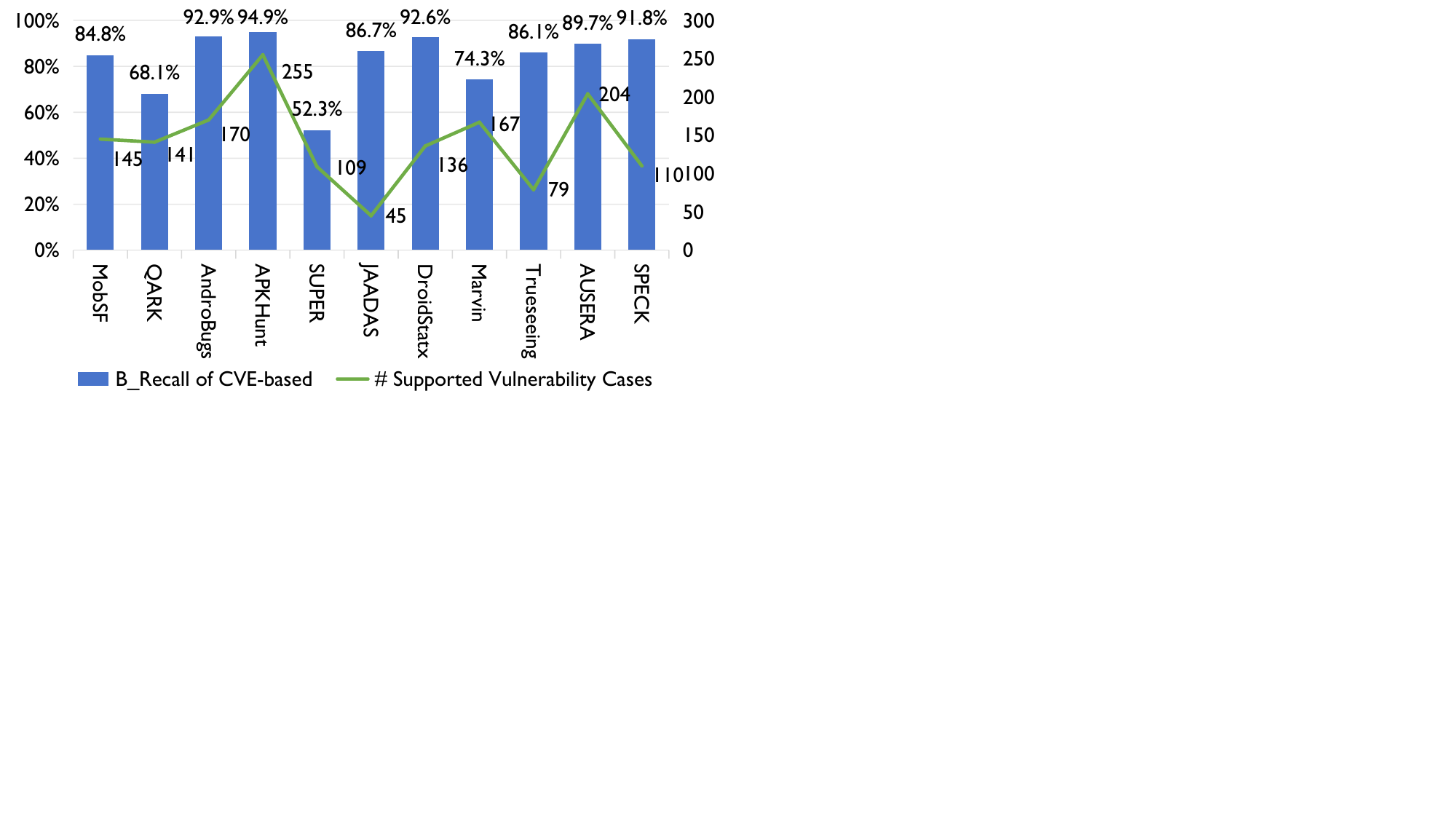}\label{fig:CVE_A_effect}}
    \subfloat[CVE-U.]{\includegraphics[width=.23\linewidth]{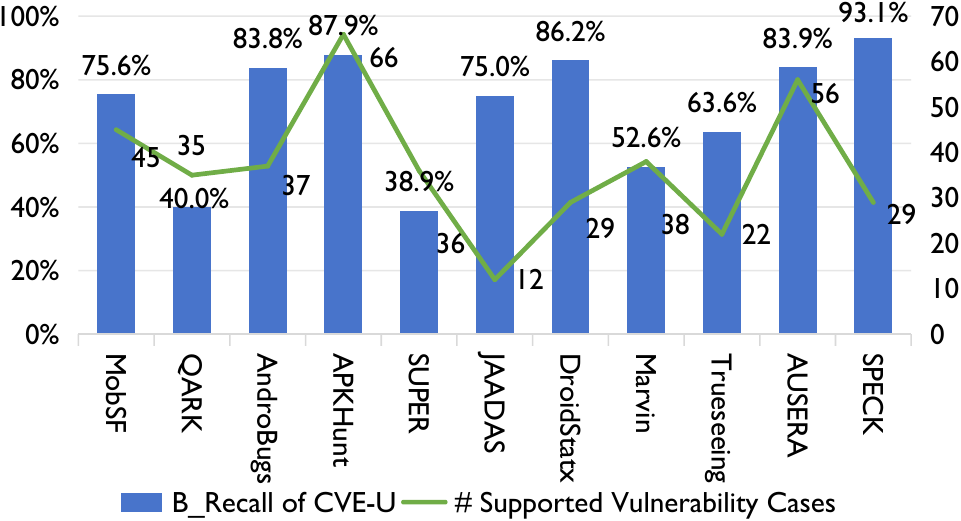}\label{fig:CVE_U_effect}}
    \caption{Effectiveness of Android SAST tools in \tool on different benchmarks. FPR refers to False Positive Rate. Since the supported vulnerability types of these SAST tools on these benchmarks are various, we highlight that the metrics shown in~\Cref{fig:tools_effectiveness} cannot be used for comparison of relative abilities across different tools, but their absolute values illustrate the detection capability on specific types of different benchmarks.}
    \label{fig:tools_effectiveness}

\end{figure*}

\subsubsection{\textbf{Result}}
\Cref{fig:tools_effectiveness} shows 11 tools exhibit broad abilities in vulnerability detection, yet many underperform expectations. 

\paragraph{\textbf{Effectiveness on the GHERA benchmark}}\label{sec:GHERA_effect}
As shown in \Cref{fig:GHERA_effect}, we marked the F1-score for all tools on GHERA along with the number of supported vulnerability cases.
{We found tools show varied effectiveness on GHERA to balance Precision and Recall.}
{For example, AUSERA's highest F1-score of 82.6\%, bolstered by a 90\% Recall, shows strong true positive (TP) identification. Despite 76\% Precision, indicating some false positives (FP) {with a 28.6\% FPR}, it balances Precision and Recall well, highlighting its effective detection.
However, certain tools have struggled to balance Recall and Precision, leading to a poor F1-score. For instance, QARK only achieved a Recall of 30.8\%, indicating missing many TPs and leading to an F1-score of just 42.1\%. 
APKHunt got a Recall of 85.7\%, covering the majority of TP. But it obtained a Precision at a mere 55.8\%, suggesting a high rate of {67.9\% FPR}. Similarly, SUPER shows an imbalance with a Precision of 66.7\% and a Recall of just 33.3\%, leading to an F1-score of only 44.4\%.}
{The remaining tools achieve a more balanced effectiveness, leading to medium-level results. For example, AndroBugs attained a 75.9\% F1-score (78.6\% Precision and 73.3\% Recall) and JAADAS exhibited a 70.6\% F1-score (75\% Precision and 66.7\% Recall).}
{In general, most tools show much underreporting on GHERA, resulting in an F1-score of no more than 85\%.}

\paragraph{\textbf{Effectiveness on the MSTG\&PIVAA benchmark}}
As shown in~\Cref{fig:MSTG_effect}, JAADAS, DroidStatx, and Trueseeing achieved 100\% B\_Recall, indicating these tools have effective detection capabilities for the types supported by this benchmark. 
Additionally, all tools, except Marvin and QARK, achieved more than 75\% B\_Recall, underscoring their effectiveness. 
This can be attributed to the vulnerabilities injected in this benchmark exhibiting simpler patterns compared to GHERA for the same type.
{
Furthermore, compared to other tools, Marvin and QARK showed lower B\_Recall of 35.7\% and 33.3\% respectively, indicating potential limitations in their detection methods when identifying types of MSTG\&PIVAA. 
Overall, most tools validated the utility of simple vulnerability patterns on this benchmark.}

\paragraph{\textbf{Effectiveness on the CVE benchmarks}}
Generally, there is no significant difference in the performance of these tools between synthetic and real-world benchmarks refer to~\Cref{fig:tools_effectiveness}. 
{For example, AUSERA achieved a B\_Recall of 89.7\% on CVE-based and 90.5\% on GHERA.}
This can be attributed to the fact that these tools rely heavily on pattern matching, detecting vulnerabilities based on the usage of specific sensitive APIs that are easy to find. This approach does not involve complex contextual analysis or cross-function examination. 
As both the selected synthetic and real-world benchmarks mainly consist of vulnerability types that are identified by the presence of certain patterns in the usage of sensitive APIs, this consistency makes the effectiveness of these benchmarks not much different.

\begin{figure*}
	\centering
	\subfloat[\footnotesize{Hardcoded Sensitive Data Exposure.}]{\includegraphics[width=.24\linewidth]{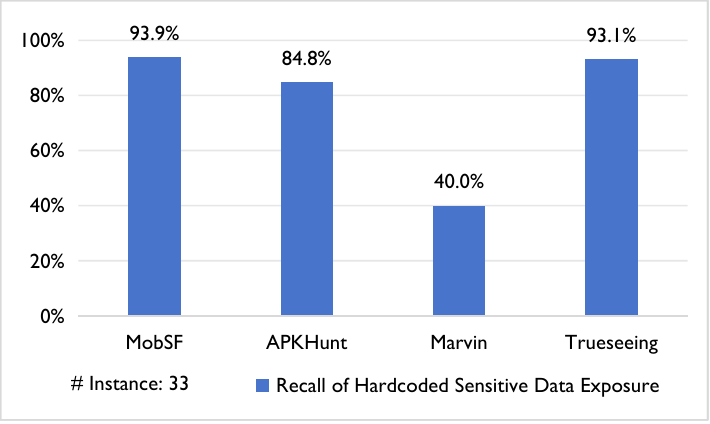}\label{fig:single_1}}\hspace{1pt}
	\subfloat[\footnotesize{Use Invalid Hostname Verification.}]{\includegraphics[width=.24\linewidth]{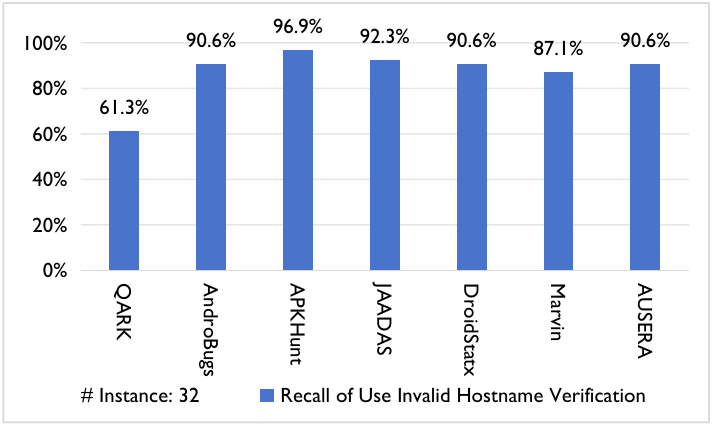}\label{fig:single_2}}\hspace{1pt}
	\subfloat[\footnotesize{Use Invalid Server Verification.}]{\includegraphics[width=.24\linewidth]{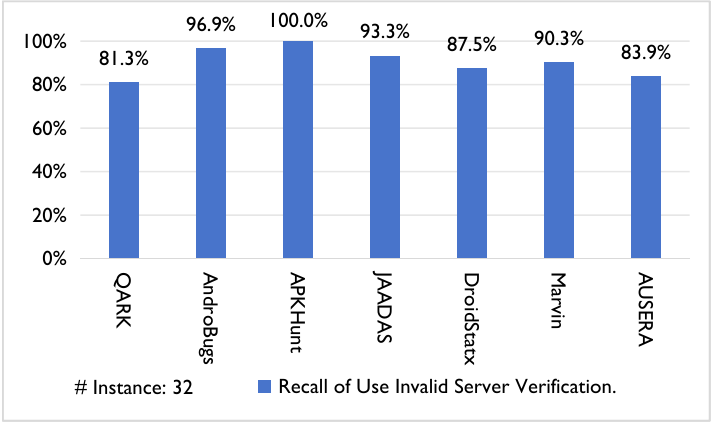}\label{fig:single_3}}\hspace{1pt}
	\subfloat[\footnotesize{Use Allow All Hostname Verification.}]{\includegraphics[width=.24\linewidth]{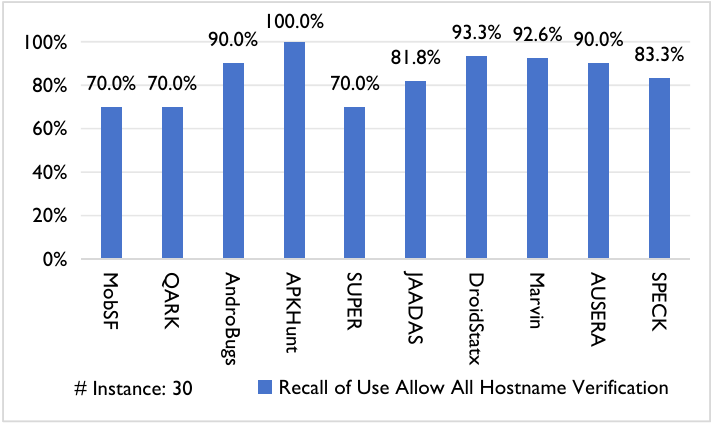}\label{fig:single_4}}\hspace{1pt}
	\subfloat[\footnotesize{Using HTTP Issue.}]{\includegraphics[width=.24\linewidth]
    {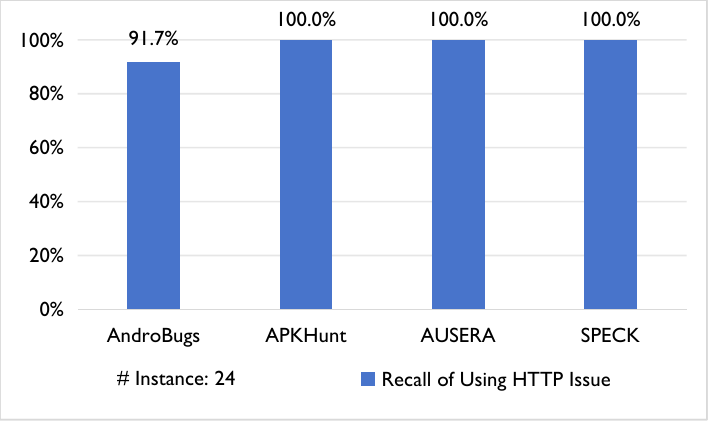}\label{fig:single_5}}\hspace{1pt}
	\subfloat[\footnotesize{Exported Not Protected Components.}]{\includegraphics[width=.24\linewidth]{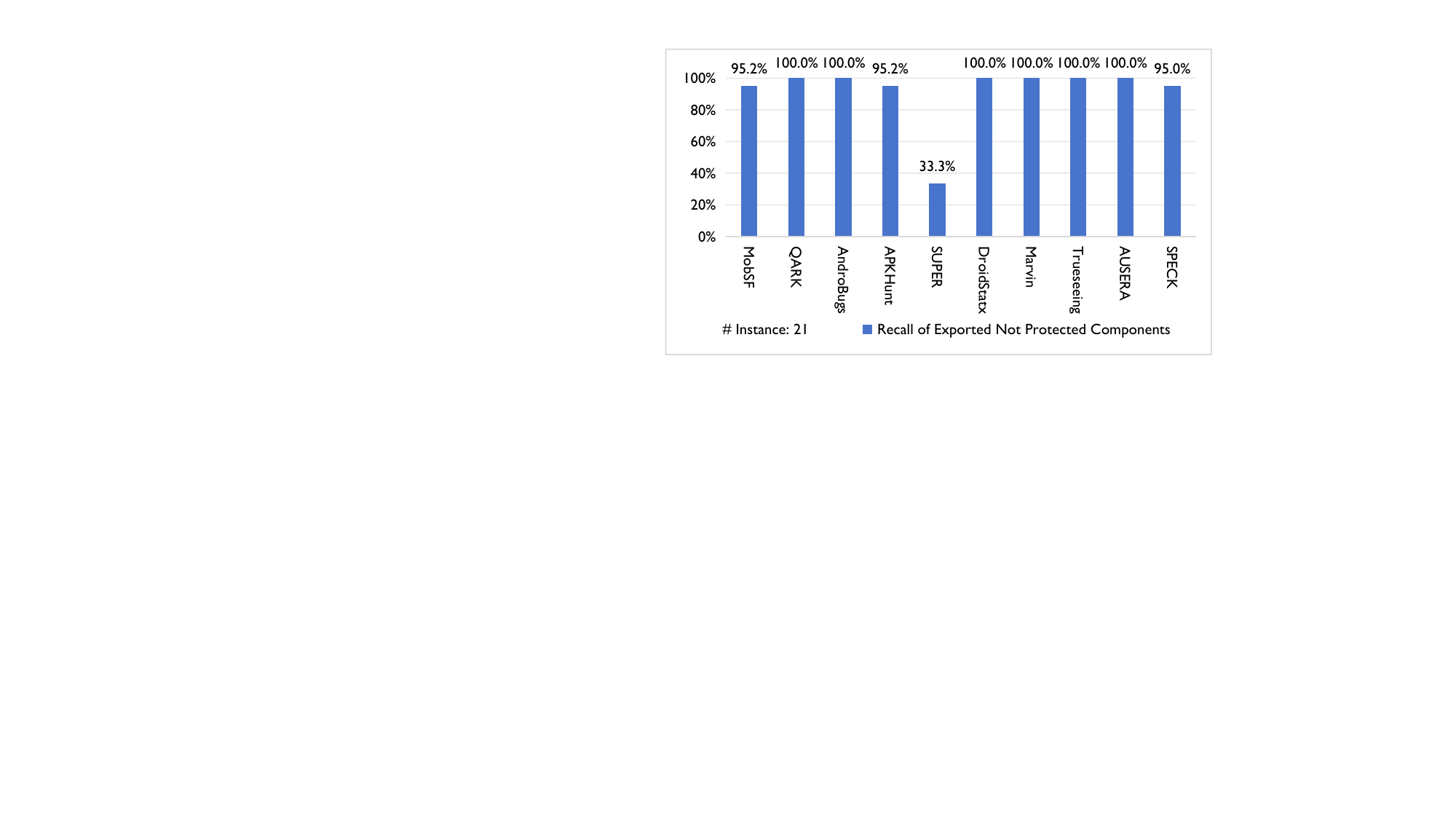}\label{fig:single_6}}\hspace{1pt}
	\subfloat[\footnotesize{Logging Data Exposure.}]{\includegraphics[width=.24\linewidth]
    {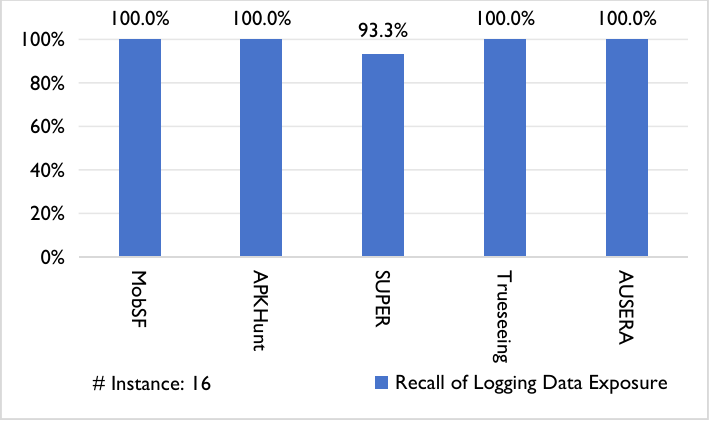}\label{fig:single_7}}\hspace{1pt} 
	\subfloat[\footnotesize{SQL Injection.}]{\includegraphics[width=.24\linewidth]
    {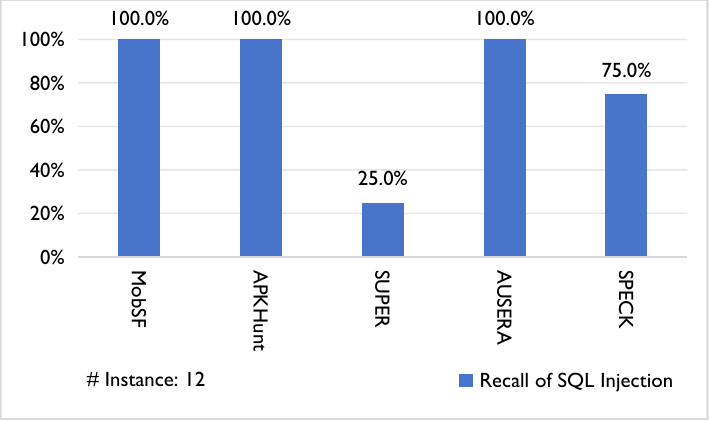}\label{fig:single_8}}\hspace{1pt}
	\subfloat[\footnotesize{External/Internal Data Exposure.}]{\includegraphics[width=.24\linewidth]{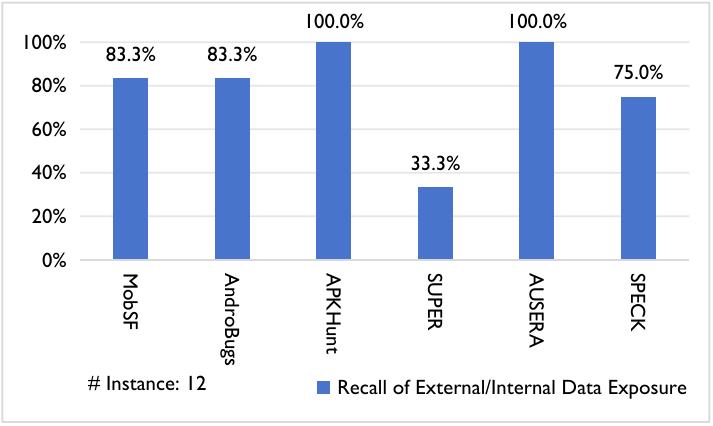}\label{fig:single_9}}\hspace{1pt}
	\subfloat[\footnotesize{Misuse Implicit Intent Issue.}]{\includegraphics[width=.24\linewidth]{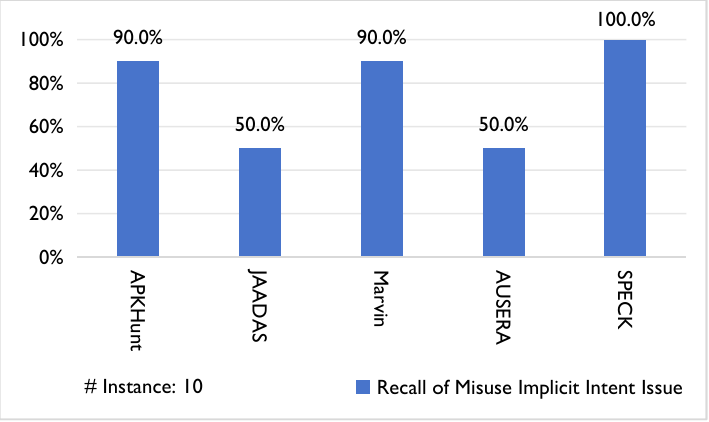}\label{fig:single_10}}\hspace{1pt}
	\subfloat[\footnotesize{Webview Local File Access.}]{\includegraphics[width=.24\linewidth]{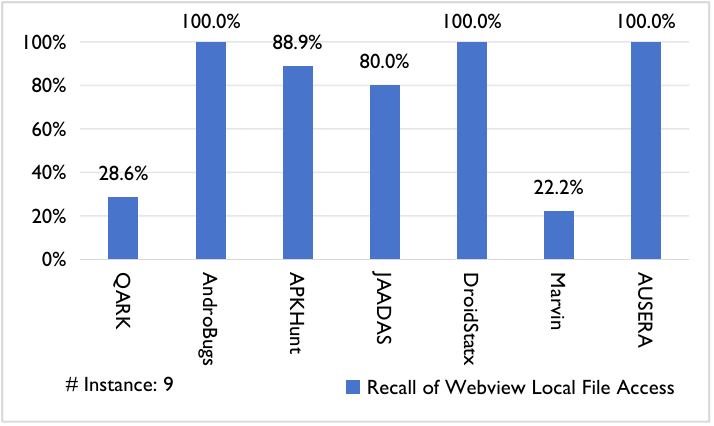}\label{fig:single_11}}\hspace{1pt}
	\subfloat[\footnotesize{Webview JavaScript Execution.}]{\includegraphics[width=.24\linewidth]{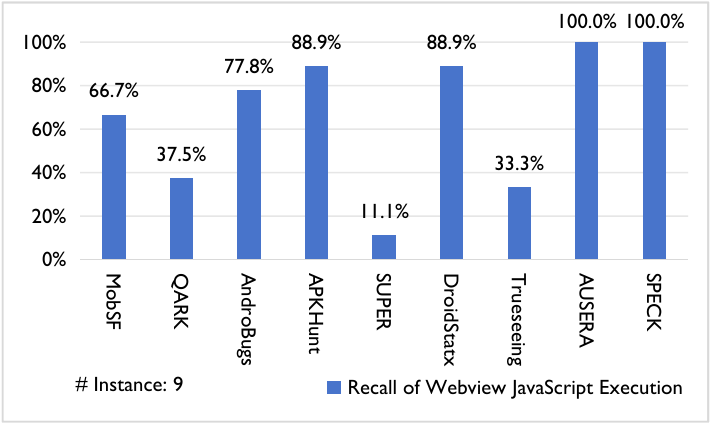}\label{fig:single_12}}\hspace{1pt}
    \caption{Effectiveness of Android SAST tools in detecting specific vulnerability types that appear over five times across the three benchmarks.}
    \label{fig:single_effectiveness}
\end{figure*}

Refer to~\Cref{fig:CVE_A_effect,fig:CVE_U_effect}, {The tools exhibit a consistent performance trend across both CVE-based and CVE-U benchmarks, with top-performing tools showing high B\_Recall in both benchmarks, while underperforming tools display low B\_Recall across them.}
SPECK achieved high B\_Recall at 91.8\% in CVE-based and 93.1\% in CVE-U. APKHunt and DroidStatx followed closely with B\_Recall at 94.9\% and 92.6\% in CVE-based and 87.9\% and 86.2\% in CVE-U, respectively. Notably, SUPER had the lowest B\_Recall both at 52.3\% in CVE-based and 38.9\% in CVE-U, with QARK performing slightly better at 68.1\% in CVE-based and 40\% in CVE-U.
Due to space limitations, we present only the top 5 CVE-based vulnerability types with the most instances in~\Cref{tab:instance num}. 
Full instance numbers per type in CVE-based are available in GitHub~\cite{androida69:online}.
Comparing the B\_Recall of CVE-based with CVE-U, we found that all tools except SPECK exhibited a marked improvement. 
{As shown in~\Cref{fig:single_effectiveness}, \Cref{fig:CVE_A_effect} and~\Cref{tab:instance num},
we observed that types that frequently occur in CVE-based and are effectively detected by tools contribute to the overall improved performance across CVE-based. 
For example, ``Hardcoded Sensitive Data Exposure'' has the most instances (33), as referred to~\Cref{tab:instance num}. Across the four supported tools in~\Cref{fig:single_1} and \Cref{fig:CVE_A_effect}, those with stronger detection capabilities in this type perform better in CVE-based, while Marvin lags due to weaker detection.
Since CVE-based reflects real-world vulnerability distribution to some extent, high detection performance on these frequent types implies the tool's effectiveness in real-world applications, suggesting that tools should focus more on detecting these frequent vulnerability types.}

\begin{table}[]
\caption{Top five vulnerability types with the most instances in CVE-based and CVE-U.}
\label{tab:instance num}
\centering
\scalebox{1}{
\begin{tabular}{@{}lrr@{}}
\toprule
\textbf{Vulnerability Types} & \textbf{\# Instance in CVE-based}  \\ \midrule
\textbf{Hardcoded Sensitive Data Exposure} & 33  \\
\textbf{Use Invalid Server Verification} & 32  \\
\textbf{Use Invalid Hostname Verification} & 32  \\
\textbf{Use Allow All Hostname Verification} & 30  \\
\textbf{Using HTTP Issue} & 24  \\ \bottomrule
\end{tabular}}
\end{table}

\paragraph{\textbf{Effectiveness on single vulnerability types}} \label{sec: effect_d}
Note that this discussion is based solely on tools' B\_Recall, as their Precision was unable to be calculated due to the nature of our benchmarks.
As shown in~\Cref{fig:single_effectiveness}, most tools generally perform well in detecting various supported types, especially for ``Logging Data Exposure'', where all tools score over 90\% B\_Recall. 
However, there are notable variations in their performance regarding specific types. For example, SUPER shows poor performance at 25\% B\_Recall for ``SQL Injection'' while the other four tools achieved a B\_Recall of at least 75\%.

{We further analyze tools' performance against specific types ordered by instance frequency, for a granular insight.}
For ``Hardcoded Sensitive Data Exposure'' in~\Cref{fig:single_1}, MobSF excels at B\_Recall of 93.9\%, closely followed by Trueseeing and APKHunt in B\_Recall of 93.1\% and 84.8\%. MobSF's superiority arises from its ability to search for hardcoded sensitive data like ``passwd'' in both source code and string pools within ``string.xml'' files. 
Trueseeing achieves high efficacy through database storage for control and data flow analysis to identify sensitive values based on characteristics like entropy and length.
However, Marvin performs poorly in this type as B\_Recall of 40\%, primarily due to its narrow focus on specific scenarios of sensitive data, such as passwords for services like Twitter and Apache credentials, rather than offering broader coverage.
Regarding ``Use Invalid Server Verification'' depicted in~\Cref{fig:single_3}, APKHunt, AndroBugs, and JADDAS showed high B\_Recall (100\%, 96.9\%, 93.3\% respectively).
{APKHunt achieved such performance by relying on rough regular expression matching for decompiled source text, whereas JAADAS and AndroBugs achieved such performance by applying a combination of data and control flow analysis built upon bytecode parsing.}
QARK's B\_Recall is only 81.3\%, largely due to incomplete decompilation of source code, exemplified by numerous empty or meaningless variables in the decompiled output of a real APK.\footnote{350apkPure.apk in CVE-based.} 
Marvin and DroidStatx missed the empty \texttt{checkServerTrusted} method in \texttt{X509TrustManager} due to reliance on bytecode analysis with strict adherence to the pattern ``\texttt{public checkServerTrusted(...)}'' and oversight of the ``\texttt{public final}'' modifier, resulting in the low performance as 90.3\% and 83.9\% respectively.

Regarding ``Using HTTP Issue'' in ~\Cref{fig:single_5}, all tools except AndroBugs demonstrate high B\_Recall (100\%). AndroBugs potentially lacks tailored detection for constructing HTTP connections via string concatenation with \texttt{HttpURLConnection}, particularly when URLs are created directly with \textit{http://}. Instead, it focuses more on instances explicitly utilizing \texttt{HttpHost} classes for HTTP connection establishment, resulting in the suboptimal performance of 91.7\% B\_Recall.
{\Cref{fig:single_6} displays that except for SUPER, all supported tools achieve high performance (over 95\% B\_Recall) in detecting ``Exported not Protected Components'' due to the straightforward detection logic, employing pattern matching for explicit or implicit exported components in the ``AndroidManifest.xml''. With 33.3\% B\_Recall, SUPER's inefficiency stems from extensive content loss during manifest file decompilation rather than flawed detection rules. SPECK misses a few cases because it focuses on broadcast and service detections, lacking activity checks, resulting B\_Recall of 95\%.}
In~\Cref{fig:single_7}, MobSF, APKHunt, AUSERA, and Trueseeing effectively detect ``Logging Data Exposure'' as all achieve 100\% B\_Recall, the strategies they employ to achieve low false negative rates differ. 
{While APKHunt and MobSF employ more loose rules by identifying sensitive API calls, like \texttt{log.e()} in decompiled source code, without validating data sensitivity or tracing its origin back to UI input. In contrast, AUSERA and Trueseeing use data flow analysis to confirm sensitive information, with AUSERA improving precision by tagging specific sensitive identifiers. 
As for SUPER, which achieved 93.3\%, it uses regular expression matching on decompiled code to search sensitive APIs involved in logging, but its reliance on hardcoded rulesets may miss variable-type sensitive data.}
In ``SQL Injection'' refer to~\Cref{fig:single_8}, limited performance (75\%) of SPECK is due to its focus on ContentProvider SQL injection, omitting SQLite and other contexts.
SUPER underperforms due to narrow pattern matching without contextual consideration leading to many FNs. As shown in~\Cref{sql_code}, to detect SQL injection, SUPER uses regex displayed at Line 2 to match. 
{However, as shown in Lines 3-16 from~\Cref{sql_code},\footnote{The vulnerable code is from ``SQLite-execSQL-Lean-benign'' in GHERA.} the query string in vulnerable code is constructed by concatenating user input and passing it to the parameter \texttt{query}. But SUPER just detects operations involving string concatenation, causing FNs.}
As displayed in~\Cref{fig:single_11}, QARK and Marvin miss many cases related to ``WebView Local File Access'', resulting in a B\_Recall of 28.6\% and 22.2\% respectively. 
Although Marvin conducts fine-grained checks by validating sensitive API-involved exported activities and analyzing exposure surfaces. 
Its overly strict rule implementation requires browsing the file scheme in the export activity, leading to severe false negatives.
Due to space constraints, we provide detailed analysis for the remaining types on GitHub~\cite{androida69:online}.

\paragraph{\textbf{Technical reasons underlying their effectiveness}}
{To ensure accuracy and reliability, two authors independently conducted the analysis, mediated disagreements with a third author, and the entire team reviewed the final results for consistency.}
{As mentioned in \S~\ref{sec:tool_selection}, all selected tools use pattern matching as core techniques. 
Therefore, their effectiveness relies heavily on hard-coded patterns, making it hard to capture vulnerable behaviors precisely. 
Specifically, coarse-grained pattern definitions boost B\_Recall but invite false positives (FP); overly fine precision increases the risk of false negatives (FN). An abundance of patterns for the same types reduces misses but escalates FPs, while overly narrow definitions lead to substantial FNs.
Applied to the selected tools, APKHunt excels in four benchmarks by using simple regular-expression matching on decompiled source code, with coarse-grained and abundant patterns leading to high B\_Recall but also many FPs in GHERA. 
For example, when detecting ``WebView JavaScript Execution'', it tries to match the presence of \texttt{setJavaScriptEnabled} API and the string \textit{WebView}, which is not enough since the vulnerability is only triggered if the API parameter is set to \texttt{true}.
Furthermore, the poor effectiveness of QARK in all four benchmarks is influenced by its limited and narrow-defined pattern matching.}
{Moreover, the overly fine-grained pattern defined leads to low B\_Recall, evidenced by Marvin for the type of ``WebView Local File Access'' mentioned earlier.}

{Well-defined detection patterns are equally critical, evidenced by the varying detection logics employed by different tools for the same vulnerability types analyzed in the above paragraph. 
For example, regarding Marvin's lower B\_Recall (35.7\%) on MSTG\&PIVAA, we analyzed its false negative cases and discovered that Marvin employed an ineffective method {to detect certain types.}
Specifically, when detecting the ``Manifest Backup Issue'', Marvin attempted to extract the ``allowBackup'' element's value in \textit{AndroidManifest.xml}. 
It flags the vulnerability if the value was set to \texttt{true}. However, in practice, it mistakenly used \texttt{android:allowBackup}, consistently extracting \texttt{None} as the value, emphasizing the importance of testing.
We displayed the original detection code alongside our corresponding fixed code in Lines 2-3 and Lines 6-7 from~\Cref{backup_code} respectively.
Also, we have identified a limitation among source-code analysis tools.
Taking APKHunt as an example, it attempts to detect the ``Mode World Storage Writable Issue'' by directly matching the string \texttt{MODE\_WORLD\_WRITEABLE}. 
This approach often results in a significant number of FNs as it relies solely on string matching without considering the subtleties of constant value resolution in decompilation.
As shown in the decompiled vulnerable source code in~\Cref{world_write_code}, the string \texttt{MODE\_WORLD\_WRITEABLE} in the original Android source code (Lines 2-4), representing the permission flag value ``2'', is decompiled to the numeric parameter ``2'' in Lines 6-8. 
This disparity underscores a key challenge: relying solely on string matching without accounting for the nuances of constant value resolution diminishes the effectiveness of source code analysis in SAST tools.
}
{Pattern matching is also constrained by its inherent limits, as it locks onto fixed vulnerability patterns, disregarding contextual consideration.}
{For instance, as previously discussed in~\S~\ref{sec: effect_d}, SUPER handles ``SQL Injection'' without context consideration.}

{Moreover, the dependency of third-party reverse or parsing tools also impacts the overall performance of the selected tools.
}
{For instance, QARK struggles with parsing certain Java files due to its reliance on the underperforming library plyj~\cite{dabeazpl40:online}, a Java syntax analysis library.
This limitation is evident when QARK fails to parse the ``NewPassword.java'' file,\footnote{A part of GHERA's BlockCipher-ECB-InformationExposure-Lean-benign.} leading to a false negative (FN), especially notable in failing to detect insecure API usages, like \texttt{Cipher.getInstance(``AES/ECB'')}. 
Moreover, Marvin incurred many FNs in MSTG\&PIVAA (6/9) due to triggered parsing errors within the SAAF framework. 
}

\begin{lstlisting}[caption={The code example of ``Mode World Storage Writable Issue''.}, label=world_write_code, float=t]
// The vulnerable source code
private void openFileOutputWorldWritable(String filename) throws Exception {
    getContext().openFileOutput(filename, Context.MODE_WORLD_WRITEABLE);}
// The vulnerable decompiled source code
private void openFileOutputWorldWritable(String filename) throws Exception {
    getContext().openFileOutput(filename, 2);}
\end{lstlisting}
\begin{lstlisting}[caption={The vulnerable code and the detection logic of ``SQL Injection'' within SUPER.}, label=sql_code, float=t]
// The regular expression used in SUPER: 
// (?:rawQuery|execSQL)\\(.*\"\\s*\\+\\s*.*\\)
protected void query(db) {
    String query = "UPDATE " 
        + MyDatabase.Table1.TABLE_NAME
        + " SET " + MyDatabase.Table1.COLUMN_NAME_VALUE 
        + " = \'" + value + "\'"
        + " WHERE " + MyDatabase.Table1.COLUMN_NAME_KEY 
        + " = \'" + key + "\'";
    try {
        db.execSQL(query);
    } catch (Exception e) {
        Log.d("E", e.toString());
    } finally {
        currentSnapshotOfTable();
    }}
\end{lstlisting}

\begin{lstlisting}[caption={The detection code of ``Manifest Backup Issue'' within Marvin.}, label=backup_code, language = Python,float=t]
# The original detection code
def check_backup(self):
    return self.apk.get_element("application", "android:allowBackup") == 'true'
# The fixed detection code
def check_backup(self):
    return self.apk.get_element("application", "allowBackup") == 'true'
\end{lstlisting}

{Based on the above in-depth analysis of cases, we have summarized 5 reasons for the tools' suboptimal effectiveness.} 
\ding{172} \textbf{\textit{Granularity issues in pattern matching.}} While nearly all 11 tools use pattern matching to detect vulnerabilities, variations in granularity were observed during the analysis of tool metadata. We conducted an in-depth examination of each tool's vulnerability detection logic at the code level, combining {the aforementioned analysis on single vulnerability types}. This involves analyzing the underlying detection approaches for every vulnerability listed in \tool across different tools. 
Based on our in-depth analysis, we find that 62.68\% (42/67) of the unified vulnerability types exhibit consistent granularity with the same logic and matching of sensitive APIs across tools for the same type. 
We paid more attention to the fine-grained granularity of {rule implementations} across these tools and concluded the main cases of different granularity. 
\textit{1)} \textbf{\textit{Data flow-sensitive vulnerabilities.}} For most data disclosure types (5.97\%, 4/67) like ``Logging Data Exposure'', As mentioned earlier, differences in tracking sensitive data and defining sensitive information lead to varying performance outcomes. 
Most tools, such as SUPER, focus primarily on matching sensitive APIs, employing relatively lenient criteria that can lead to higher false positives. In a real-world APK sample, SUPER flagged 597 instances of log data exposure, whereas APKHunt reported 1,210 instances, thus increasing developers' review burden.
\textit{2)} \textbf{\textit{Vulnerabilities with preconditions.}} For vulnerabilities that only trigger with certain preconditions, we find that different tools have different detection granularity. 
For example, the sensitive API \texttt{setAllowFileAccess(`true')} in ``Webview Local File Access'' only triggers for min SDK version below 17 while only AUSERA and QARK conduct API matching with further validation of the min SDK version. 
Five vulnerability types having the constraints of preconditions are in this category, accounting for 7.46\% (5/67).
\textit{3)} \textbf{\textit{Omitted detection of certain sensitive APIs.}} For vulnerability types with multiple sensitive APIs (23.88\%, 16/67), differences arise when tools omit certain sensitive APIs in the analysis. For example, most tools only check for AES encryption misuse via \texttt{Cipher.getInstance("AES/ECB")}, while ignoring the implementation of \texttt{Cipher.getInstance ("AES")} also uses the parameters ``AES/ECB/PKCS5padding''. 
Among the 3 causes we discussed, the first two as rough detection granularity tend to yield excessive false positive results, {highlighting the need for well-tuned granularity, avoiding extremes of coarseness and fineness.}
The last may suffer from numerous false negatives due to disparities in sensitive API coverage which calls for reasonable coverage.
\ding{173} \textit{\textbf{Detection logic issues.}} As shown in the case of Marvin detecting backup issues, inconsistencies arise between the claimed detection capabilities of tools and their actual performance, often arising from issues in their detection logic.
\ding{174} \textbf{\textit{A lack of code context.}} Concerning SQL Injection, despite source code analysis tools capturing high-level language structures, their detection logic often relies on pattern matching without contextual consideration, leading to instances of false negatives.
\ding{175} \textit{\textbf{Issues with integrated tool libraries.}} Tools such as QARK relies on plyj, and Marvin relies on SAAF, any problems with the libraries they depend on can significantly impact their performance as mentioned earlier.
\ding{176} \textbf{\textit{Decompilation issues.}} Decompilation tools cannot perfectly reconstruct source code, leading to issues such as missing code snippets and parameter variations. {Just as discussed in~\S~\ref{sec: effect_d}, the decompilation content omissions would lead to false negatives when using QARK and SUPER.}

\smallskip
\noindent\fbox{
\parbox{0.95\linewidth}{
\textbf{Answer to RQ3:} 
\ding{172} {All evaluated tools exhibit suboptimal effectiveness across four benchmarks. Specifically, QARK achieved the lowest F1-score on GHERA at 42.1\% while SUPER had the lowest B\_Recall on CVE-based at 38.8\% and on CVE-U at 38.2\%. On MSTG\&PIVAA, QARK obtained the lowest B\_Recall of 33.3\%}
\ding{173} {Source code-based tools, like QARK and SUPER tend to experience effectiveness fluctuations affected by the quality of decompiled source code.}
\ding{174} {Varying degrees of detection inconsistency among tools can be found like Marvin can not detect ``Manifest Backup Issue'' as its detection code bug.}
{\ding{175} {The performance of the tools on synthetic and real-world benchmarks in our study did not differ significantly.}}
}}

\begin{figure}
\centering   
\includegraphics[width=0.45\textwidth]{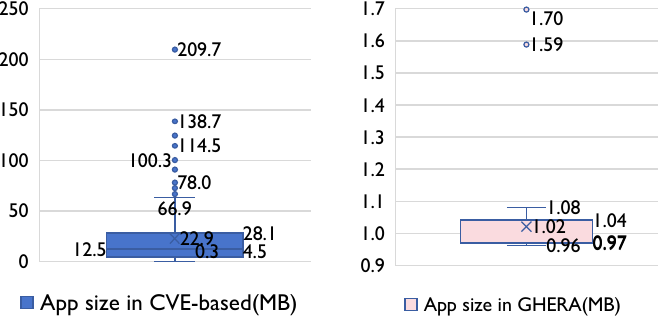}
\centering
\caption{The size distribution of the sample APK in CVE-based and GHERA.}
\label{fig: apk_size}
\end{figure}

\begin{table}
\caption{Time performance comparison of different tools.}
\label{tab:timecost}
\centering
\scalebox{0.8}{\begin{tabular}{@{}lccccccccccc@{}}
\toprule
\textbf{Tool} & \textbf{\rotatebox{90}{MobSF}} & \textbf{\rotatebox{90}{QARK}} & \textbf{\rotatebox{90}{AndroBugs}} & \textbf{\rotatebox{90}{APKHunt}} & \textbf{\rotatebox{90}{SUPER}} & \textbf{\rotatebox{90}{JAADAS}} & \textbf{\rotatebox{90}{DroidStatx}} & \textbf{\rotatebox{90}{Marvin}} & \textbf{\rotatebox{90}{Trueseeing}}  & \textbf{\rotatebox{90}{AUSERA}}  & \textbf{\rotatebox{90}{SPECK}}  \\ \midrule
\textbf{Time (s)} & 197.6 &	191.7 &	14.9 &	172.6 &	8.5 &	63.2 &	21.7 &	49.6 &	63.3 &	45.0 &	40.7 \\ 
\textbf{Failed (apk)} & 0 & 8 & 0 & 0 & 3 & 142 & 1 & 12 & 5 & 23 & 9 \\ 
\bottomrule
\end{tabular}}
\end{table}

\subsection{RQ4: Time performance} \label{sec：rq4}

\subsubsection{\textbf{Setup}}
To investigate the tools' time performance, we employed all {305} sample apps from GHERA, MSTG\&PIVAA, and CVE-based benchmark to analyze detection time, running each tool three times to avoid bias or unexpected errors,  

\subsubsection{\textbf{Result}}
{\Cref{tab:timecost} shows the average time taken by each tool for scanning a single APK.}
{We found that MobSF required the longest scanning time ({197.6s}), followed by QARK (191.7s) in the second longest position while APKHunt was the third (172.6s). }
MobSF takes longer due to its extensive detection scope. In addition to code analysis, it conducts further examination including security analysis on binary files, such as checking the NX bit status in ELF files.
{QARK takes a longer time due to its uses of three decompilers in parallel. Since the entire decompilation waits for the last one to finish, an increase in the runtime of any single decompiler will extend the overall scan time, thus slowing down the process.}
APKHunt exhibits longer analysis time as it traverses each decompiled Java file to perform detection for all supported vulnerability types as it supports more types.

We observed that SUPER exhibits the shortest time and thus the best time performance overall {for its utilization of parallel scanning, which significantly reduces the scan time.}
AndroBugs took slightly longer than SUPER for its scans. The accelerated scanning is achieved through AndroBugs' modification of Androguard, allowing based on bytecode analysis.
DroidStatx took only {21.7s} per APK, attributed to its focus on analyzing \textit{smali} files derived from dex files, a less time-consuming process than decompiling dex files into source code.
The remaining tools have a similar time cost ranging from 40s to {70s}.
{Our findings revealed that among the tools with average scanning times below 100s, 75\% (6/8) rely on bytecode-based analysis. This implies that bytecode-based tools are faster as bytecode is a less complex representation than source code, simplifying analysis. By contrast, tools employing decompilation to source code face substantial computational demands because the decompilation process itself is highly time-consuming.}

{To better understand the time performance of different tools, we further present the size of APKs in CVE-based and GHERA in~\Cref{fig: apk_size} (the app size in MSTG\&PIVAA is 5.9MB and 3.2MB respectively), and the variation in scanning time of each tool across different APK size intervals in~\Cref{fig: time_size}.}
{Refer to~\Cref{fig: apk_size}, the distribution of APK sizes within the CVE-based spans a wide range, extending from 0.3 MB to 209.7 MB, with most (75\%) beneath {27.9} MB and just a few (9) surpassing 100 MB while the APK size in GHERA is concentrated near 1MB.}
{Detection time typically rises with increasing APK size across most tools, as anticipated for larger files. Notably, APKHunt experiences a significant increase in detection time for APKs over 100MB, suggesting it takes longer to analyze larger files.}

During the tool scanning process, we identified cases in which certain tools failed to obtain scanning results.
Specifically, out of the {305} APKs across three benchmarks (considering GHERA includes both benign and secure APKs), the number of failure cases for each tool is detailed in~\Cref{tab:timecost}. Notably, JAADAS exhibited a high number of failed scans, with 142 instances.
Failure cases in JAADAS stem from analysis issues within Soot, which it relies upon. When failures are due to timeouts, the root cause lies in discernible pauses occurring during the Soot analytical procedure.
The failure cases of SUPER were due to the unsuccessful DEX to JAR conversion using Dex2Jar. 
In summary, tool failures stemmed from three main reasons:
\ding{172} Inherent flaws in the tools' scanning logic, which leads to unsuccessful scans. For instance, Marvin attempted to convert a string to an integer without accounting for the presence of the \texttt{0xa0} string.
\ding{173} Unsuccessful decompilation of APK, such as the flaw in QARK's decompile function, leading to scan terminations. 
\ding{174} Failure during analysis. In this case, bytecode analysis frameworks the SAST tools depend on such as Soot encounter failures in analyzing, specifically encountering exceptions during processing. Examples include AUSERA and JAADAS.

\smallskip
\noindent\fbox{
\parbox{0.95\linewidth}{
\textbf{Answer to RQ4:} {
{\ding{172} The bytecode-based tool (e.g., AndroBugs) scans faster than most tools that employ source code analysis. 
\ding{173} The selection of decompilers significantly influences the scanning speed of the tool. For instance, QARK employs three different decompilers, which results in an increased time cost (i.e., 175.4s) for its scanning process.
\ding{174} {These tools demonstrate varying degrees of scan failures. Notably, JAADAS experienced 59 scan failures, attributable to a bug within the Soot framework.}
}}}}

\section{Discussion}
\subsection{Implications}
\subsubsection{{Suggestions for SAST tool developers}}
To enhance Android vulnerability detection capability, we propose the following suggestions for SAST tool developers.

\noindent\textbf{(1) {Expand coverage for overlooked vulnerability types.}} 
{As discussed in \S~\ref{sec:rq1}, many tools neglect certain vulnerability types.}
In comparison to our 67 unified vulnerability types, the highest coverage is merely 67\%.
{Beyond taxonomy overlaps, our study found notable differences in the unique vulnerability types supported by tools. APKHunt leads with 15 unique types, while some tools have none.}
{Hence, tool developers can use our taxonomy and unique types list supported by each tool (shared on GitHub~\cite{androida69:online}) as a baseline for expanding supported types in their tools.}
Further, as detailed in \S~\ref{sec:rq2}, 79\% of Android-specific CVEs' unsupported vulnerability types, and 65\% of 23 types in GHERA are undetectable to rely solely on pattern matching.
{This exposes a significant gap between the detection capabilities of current SAST tools and the security needs of applications.}
Therefore, to better identify vulnerability types, developers should prioritize expanding detection capabilities for overlooked yet common types.
Exploring alternative detection techniques beyond pattern matching is essential.

\noindent\textbf{(2) Improve the effectiveness of vulnerability detection.}
In \S~\ref{sec: rq3}, we discussed five technical reasons underlying tool effectiveness. Here are some suggestions for developers:
{\ding{172} Use more detailed detection patterns to cover various vulnerability scenarios and prevent false negatives, as highlighted in~\S~\ref{sec: rq3}.}
{\ding{173} For tools that rely on decompilation tools for source code analysis, they should enhance their detection performance by incorporating code context. While not a novel tip, there are still many tools that have not implemented it.}
{\ding{174} Ensure the usability of integrated analysis frameworks, implement robust exception handling, and regularly update tools to their latest versions. For example, the analysis bugs in Soot or failed scan caused by Dex2Jar (as mentioned both in~\S~\ref{sec：rq4}).}
{\ding{175} Test and verify the claimed vulnerability detection logic to align with actual results and avoid discrepancies. For instance, despite the simplicity of Marvin's approach for detecting ``Manifest Backup Issues'', its simple bug resulted in misidentification.}

\noindent\textbf{(3) Evaluate tools on suitable benchmarks in consistency.}
{\S\ref{sec: rq3} reveals discrepancies between the vulnerability types covered by tools versus those represented in three benchmarks while the highest coverage is 88\% from APKHunt at CVE-based. Developers need appropriate benchmarks to evaluate tool performance. The open-source community also urges the creation of benchmarks that cover a broader range of types.}

\noindent{\textbf{(4) Optimize the integration of decompilers.}}
As detailed in~\S~\ref{sec：rq4}, the decompilation time greatly affects time cost because source-code analysis tools rely on decompiled source code. 
{Tool developers could evaluate the effectiveness and necessity of decompilers in vulnerability detection, and consider removing redundant or underperforming decompilers to reduce scan times.}

\subsubsection{{Suggestions for app developers}}
For better SAST tool selection for vulnerability detection, we give suggestions for app developers.

\noindent\textbf{(1) Select SAST tools via specific app security requirements.} 
According to our analysis in \S~\ref{sec:rq1}, no single tool can completely cover all the vulnerability types contained in our proposed taxonomy, indicating the importance of SAST tool selection with application-specific requirements. App developers should select SAST tools aligned with their specific requirements related to their focus on app features.
For example, when assessing apps involving sensitive data, AUSERA which provides more attention to detection of data leakage issues should be prioritized.

\noindent\textbf{(2) Select SAST tools based on the need for time cost.} 
For high-time performance needs, choose lightweight bytecode SAST tools like AndroBugs given their efficiency. In scenarios where pursuing vulnerability detection rate of detection with flexible time budgets, using multi-decompiler tools like QARK accepts higher time costs for enhanced detection.

\subsection{Threats to Validity}
\subsubsection{\textbf{External Validity}}
{An external threat involves using CVE as our sole real-world vulnerability source. This limitation potentially constrains our analysis's comprehensiveness and universality. However, our diverse, large in size, and systematically constructed CVE-based benchmark mitigates this by encompassing 34 vulnerability types and 262 instances enhancing our finding's relevance, applicability, and reliability.}
{
Another possible external threat exists from building the CVE-based benchmark. This threat is intensified by the process of labeling the filtered CVEs with vulnerability types defined in our proposed taxonomy, encompassing both the overlapped types and the unique types.}
Since some descriptions contained in CVE entries lacked clarity, we traced back to the resources linked within each entry to obtain confirmed explanations to better label. This helped mitigate potential labeling bias arising from vague descriptions. We also conducted a cross-validation approach to eliminate human bias.
Furthermore, our study focuses on evaluating Android SAST tools that detect general vulnerability types, excluding those designed for specific types. The experiments were designed to evaluate Android SAST tools that detected general vulnerability types, rendering it inappropriate to include specialized tools. Despite this limitation, our assessment of general vulnerability detection tools still offers valuable insights for the field.

\subsubsection{\textbf{Internal Validity}}
The internal threat to the effectiveness of our research comes from artificially constructed unified taxonomy. 
While we have thoroughly examined and compared the source code of each tool, potential human bias and errors during the extraction and mapping of detection rules remain a concern.
To mitigate this threat, we have refined our taxonomy via cross-validation by all authors. 
In addition, in the experiments we have done, tools are executed in their default configuration. The default configuration of different tools may not be able to fully perform their functions, which may affect their detection results. However, we limit the experiment to the default configuration, because this is the most likely configuration for most users.  

\section{Related Work}
Validating the effectiveness of Android SAST vulnerability detection tools has become an important research direction. Currently, evaluations mainly rely on synthetic benchmarks or {serveal} real-world apps.
For example, Ranganath et al.~\cite{ranganath2020free} evaluated 14 Android SAST tools 
on GHERA, a synthetic benchmark proposed by Mitra et al.~\cite{mitra2017ghera}.
{The study used GHERA's coarse-grained categories (e.g., ICC) to identify vulnerability types for tool evaluation leading to a rough correspondence between the tools' supported types and the GHERA categories, while our evaluation delved into finer-grained types, providing a more precise and detailed unified mapping of the vulnerability types each tool can detect.}
Chen et al.\cite{chen2020empirical,chen2022ausera} introduced AUSERA, a SAST tool with the capability of automated vulnerability detection for Android apps, and conducted an evaluation of 5 SAST tools.
Meanwhile, the study revealed several reasons for the false positives introduced by the tools.
Reaves et al.~\cite{reaves2016droid} conducted a systematic analysis of the literature involving Android security research, providing a comprehensive overview of Android SAST tools and a discussion of the techniques and frameworks used in Android SAST tools. In addition, they evaluated 7 SAST tools based on the tools' ease of use and successful scanning cases on a set of Google Play apps. 
Senanayake et al.~\cite{senanayake2023android} also discussed the Android vulnerability detection method based on comprehensive related literature and provided an overview of the vulnerability detection method based on machine learning and traditional methods (i.e. static analysis and dynamic analysis).

However, the research mentioned above does not take into account the inconsistency between vulnerability types supported by the evaluation tools and vulnerability types supported by the benchmark, which will introduce a certain bias in the evaluation. In other words, the comparisons can only focus on coarse-grained quantities instead of fine-grained vulnerability types. Meanwhile, evaluation only by the synthetic benchmarks is limited.
Our work proposed a unified taxonomy that contains 67 vulnerability types that can help construct a benchmark that can better match the detection capabilities of different tools, leading to more fine-grained evaluation results. Additionally, both synthetic benchmarks and real-world benchmarks have been investigated in this work.

Several prior studies have conducted evaluations of SAST tools in different contexts such as Java~\cite{fse23,ase-12,liu2023comprehensive}, JavaScript~\cite{brito2023study}, C/C++~\cite{issta22-c}, and Solidity~\cite{li2024static}. 
For instance, Li et al.~\cite{fse23} compared 7 free-of-charge SAST tools using the OWASP Benchmark and a constructed CVE Benchmark consisting of 165 unique Java CVEs.
Notably, while their findings coincide with our findings on the limitations of synthetic benchmarks, 
{our study scope, distinct from it, focuses on Android SAST tools, given the differences between the Android and Java ecosystems, such as communication mechanisms, which lead to distinct vulnerabilities. Our research delves into the technical gaps in Android SAST tool performance for detecting general vulnerabilities and conducts a quantitative analysis, emphasizing the need for systematic research to reveal insights in the Android domain, separate from the Java domain.}

In summary, our work distinctively contributes to the state of the art through the following aspects:
\textit{\textbf{1)}} \textbf{Target domain} (focused on general Android SAST tools), 
\textit{\textbf{2)}} \textbf{Benchmarks used} (use of synthetic benchmarks and CVE-based benchmark),
\textit{\textbf{3)}} \textbf{Evaluation methodology} (introduction of a unified vulnerability taxonomy plus a scalable and automated evaluation platform (\tool)), and
\textit{\textbf{4)}} \textbf{Evaluation scope} (inclusive of aspects like vulnerability type coverage and consistency, detection effectiveness, and time performance).

\section{Conclusion}
In this paper, we have taken the first step to build a unified platform \tool, which contains 67 general/common vulnerability types and is further used to comprehensively and effectively evaluate Android SAST tools. 
We then evaluated 11 selected Android SAST tools on both {our newly constructed} real-world benchmarks and existing synthetic benchmarks.
{Our study reveals numerous valuable insights into the tools' performance and
provides clear guidance for future optimization and improvement of the tools and an innovative perspective to complement previous work analyzing SAST tools. 
Future work can focus on developing a more effective and efficient tool based on the insights gained from this paper.}

\section*{Acknowledgements}
We thank the reviewers for their insightful comments.
This work was supported by the National Natural Science Foundation of China (No. 62472309, 62102283), and the Natural Science Foundation of Tianjin (No. 22JCYBJC01010).

\bibliographystyle{IEEEtran}
\bibliography{reference}

\begin{IEEEbiography}[{\includegraphics[width=1in,height=1.25in,clip,keepaspectratio]{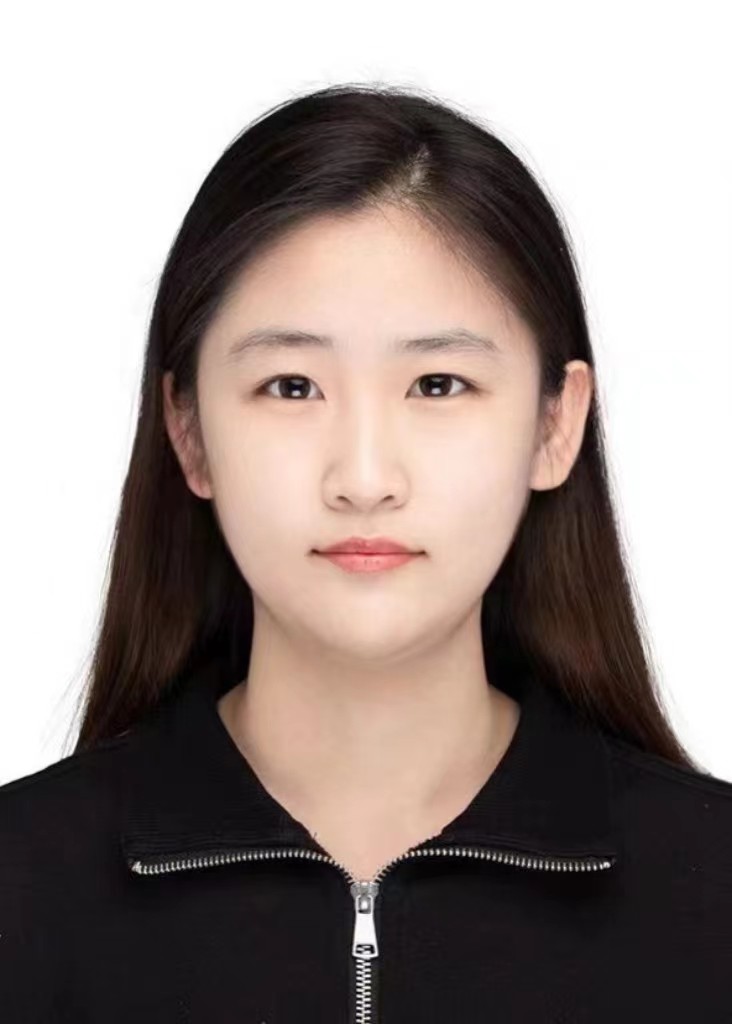}}]
{Jingyun Zhu} is a master student at the College of Intelligence and Computing, Tianjin University, China. Her research interests focus on Android vulnerability detection and static analysis.
\end{IEEEbiography}

\begin{IEEEbiography}[{\includegraphics[width=1in,height=1.25in,clip,keepaspectratio]{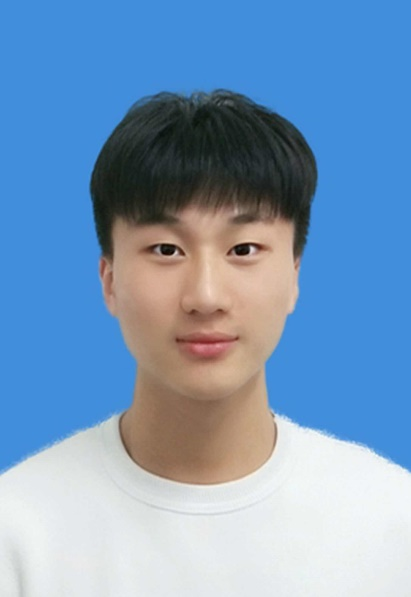}}]
{Kaixuan Li} (Member, IEEE) is currently a Ph.D. student at the Software Engineering Institute, East China Normal University, China, and a research assistant at Nanyang Technological University, Singapore. His research focuses on static analysis and large language models. He got an ACM SIGSOFT Distinguished Paper Award at FSE 2024. More information is available on {\url{https://kaixuanli-ecnu.github.io/}}.
\end{IEEEbiography}

\begin{IEEEbiography}[{\includegraphics[width=1in,height=1.25in,clip,keepaspectratio]{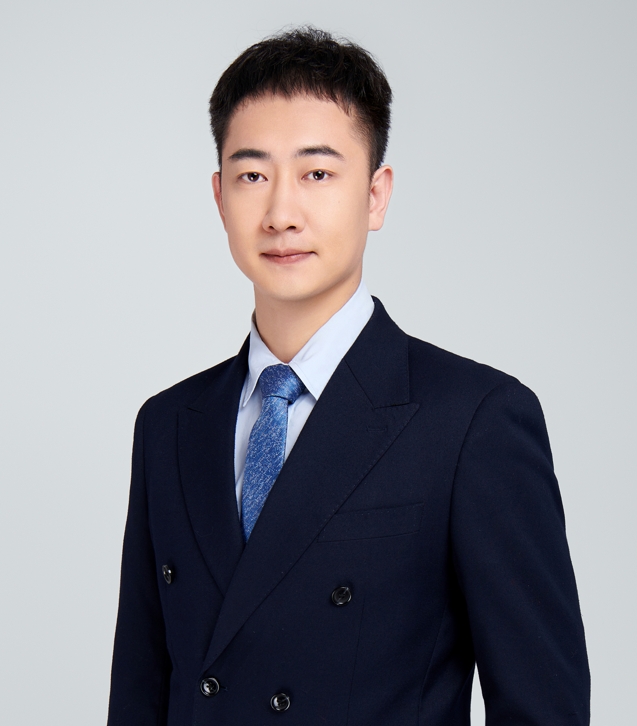}}]
{Sen Chen} (Member, IEEE) is an Associate Professor at the College of Intelligence and Computing, Tianjin University, China. 
Before that, he was a Research Assistant Professor at Nanyang Technological University, Singapore.
His research focuses on software and system security.
He got 6 ACM SIGSOFT Distinguished Paper Awards. More information is available on {\url{https://sen-chen.github.io/}.}
\end{IEEEbiography}

\begin{IEEEbiography}[{\includegraphics[width=1in,height=1.25in,clip,keepaspectratio]{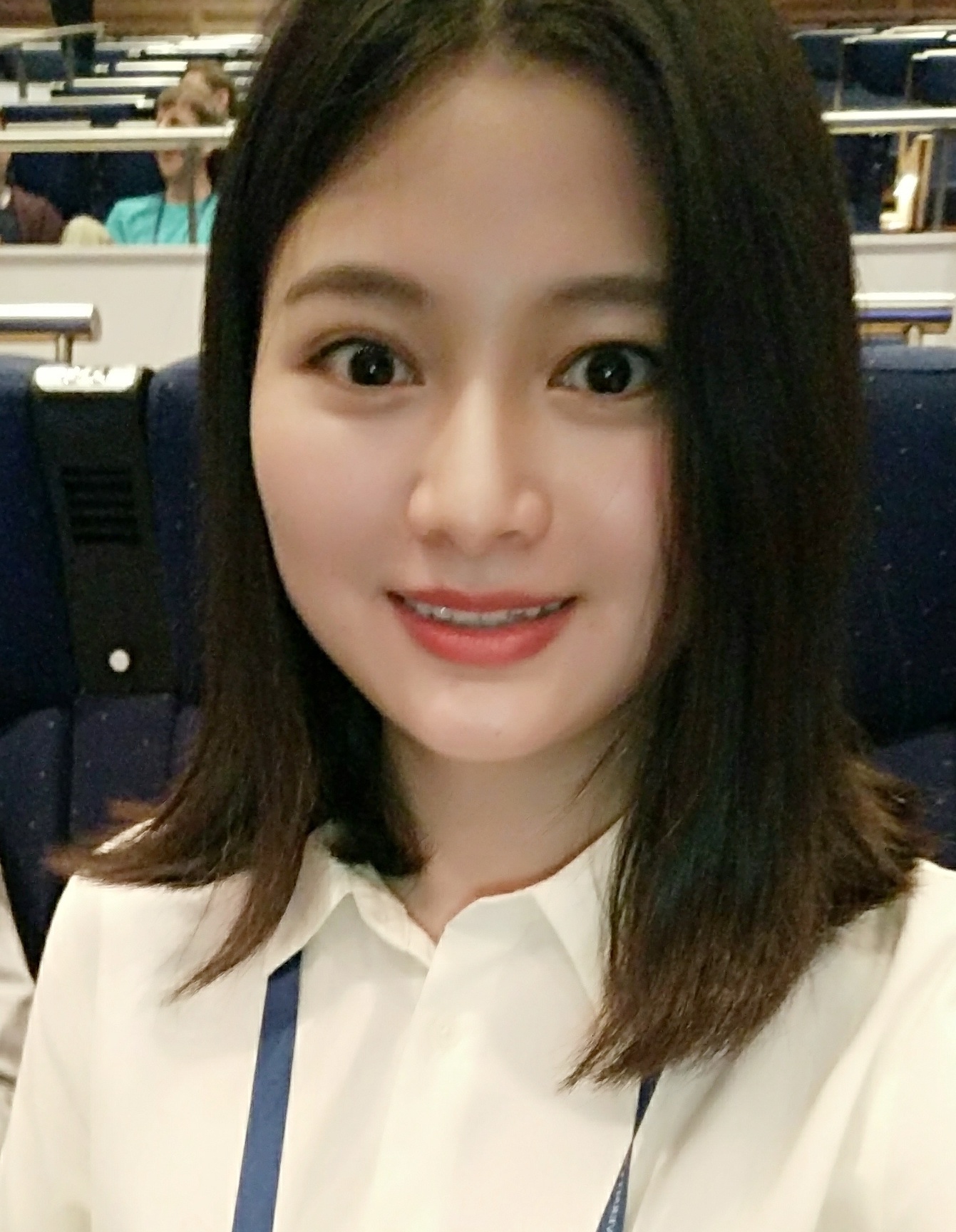}}]
{Lingling Fan} is an Associate Professor at the College of Cyber Science, Nankai University, China. 
In 2017, she joined Nanyang Technological University (NTU), Singapore as a Research Assistant and then had been a Research Fellow of NTU since 2019. Her research focuses on program analysis and testing, and software security. She got 4 ACM SIGSOFT Distinguished Paper Awards at ICSE 2018, ICSE 2021, ASE 2022, ICSE 2023.
\end{IEEEbiography}

\begin{IEEEbiography}[{\includegraphics[width=1in,height=1.25in,clip,keepaspectratio]{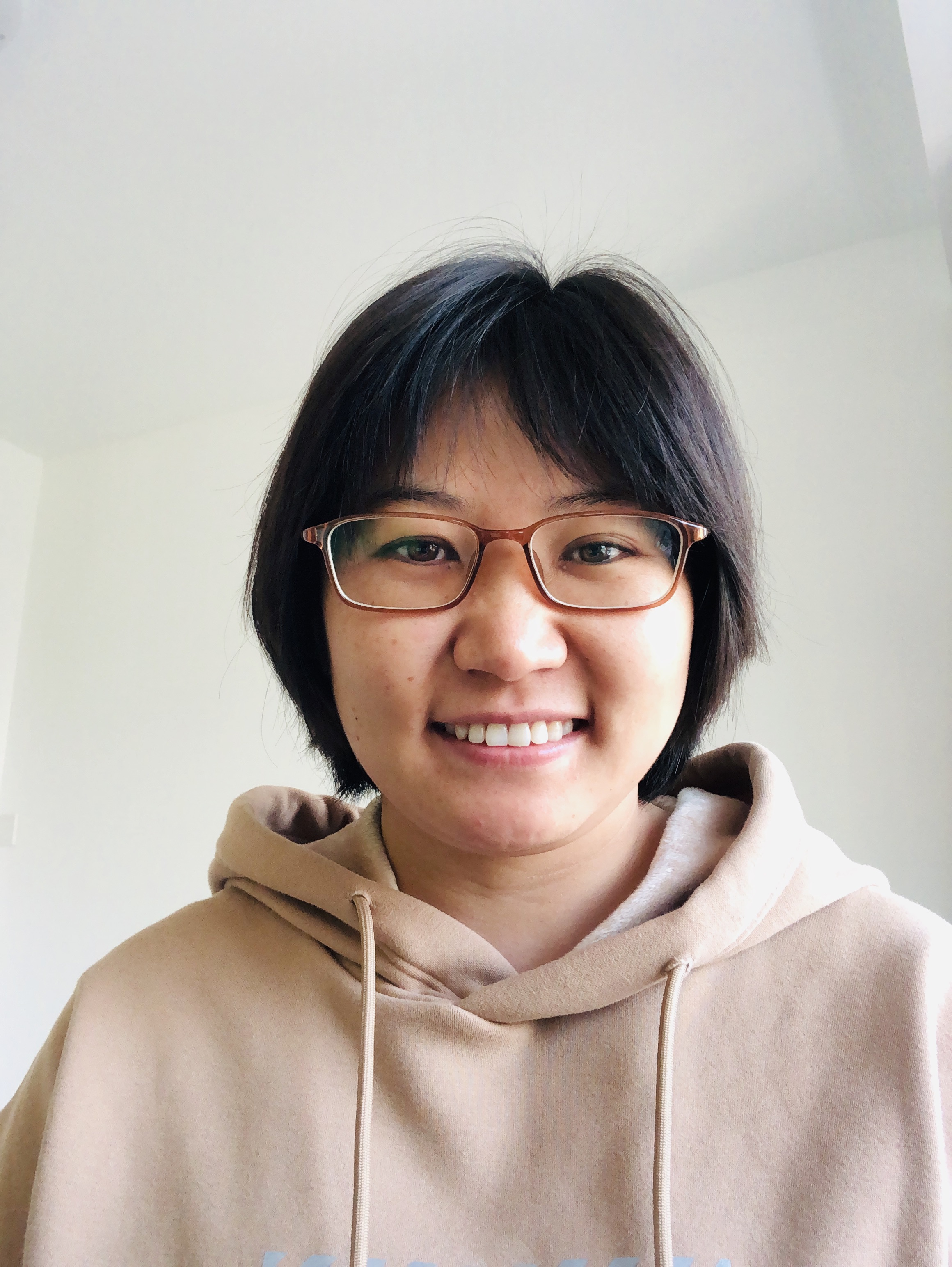}}]
{Junjie Wang} is an Associate Professor at the College of Intelligence and Computing, Tianjin University, China. 
Before that, she was graduated from the School of Computer Science and Engineering, Nanyang Technological University, Singapore. Her research focuses on vulnerability detection. She found dozens of vulnerabilities in widely used products of Microsoft, Apple, Google and won the title of MSRC most valuable security researcher. More information is available on {\url{https://zhunki.github.io/}.}
\end{IEEEbiography}

\begin{IEEEbiography}[{\includegraphics[width=1in,height=1.25in,clip,keepaspectratio]{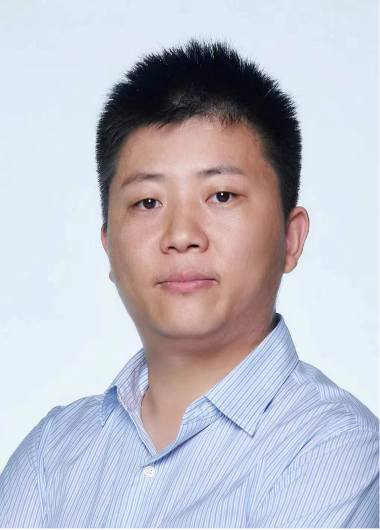}}]
{Dr.Xiaofei Xie} is an Assistant Professor and  Lee Kong Chian Fellow at Singapore Management University. He obtained his Ph.D from Tianjin University and won the CCF Outstanding Doctoral Dissertation Award (2019) in China. Previously, he was a Wallenberg-NTU Presidential Postdoctoral Fellow at NTU. His research mainly focuses on the quality assurance of both traditional software and AI-enabled software. He has published top-tier conference/journal papers in the areas of software engineering, security and AI, focusing on the use of AI for software testing and the testing and security of AI systems. In particular, he has received four ACM SIGSOFT Distinguished Paper Awards (FSE'16, ASE '19, ISSTA '22 and ASE '23) and a APSEC Best Paper Award.
\end{IEEEbiography}

\vfill

\end{document}